\documentclass[english,3p,twocolumn,authoryear]{elsarticle}

\usepackage[T1]{fontenc}
\usepackage[latin1]{inputenc}
\usepackage{amsmath}
\usepackage{setspace}
\usepackage{amssymb}
\usepackage{color}
\makeatletter
\usepackage{setspace}
\usepackage{epic,epsfig}
\usepackage{subfigure,float}
\usepackage{amsmath,amsfonts,amssymb}
\usepackage{inputenc,euscript}
\usepackage{graphicx,natbib,amssymb,lineno}
\usepackage{tabularx,enumerate}
\usepackage{lscape}
\usepackage{rotating}
\usepackage{verbatim}
\usepackage{subfigure}
\makeatother

\newcommand{\kave}{$\bar{k} $ } 
\newcommand{\kaven}{$\bar{k}$}  % no space after the command
\newcommand{\cave}{$\bar{c} $ } 
\newcommand{\caven}{$\bar{c}$}  % no space after the command

\newcommand{\lave}{$\bar{l} $ }
\newcommand{\laven}{$\bar{l}$}  % no space after the command
\newcommand{\knnave}{$\overline{k_{nn}} $ }
\newcommand{\nl}{$N_{LC} $ } 
\newcommand{\nlave}{$\overline{N_{LC}} $ }
\newcommand{\nlaven}{$\overline{N_{LC}}$}  % no space after the command
\newcommand{\LFF}{\emph{lastfm} }
\newcommand{\LFFn}{\emph{lastfm}}  % no space after the command
\newcommand{\email}{\emph{email} }
\newcommand{\emailn}{\emph{email}}  % no space after the command

\newcommand{\TFCn}{\emph{TCG}}  % no space after the command
\newcommand{\tfc}{triadic closure and global connections }
\newcommand{\tfcn}{triadic closure and global connections}
\newcommand{\vaz}{V\'az }
\newcommand{\vazn}{V\'az}  % no space after the command
 \newcommand{\er}{Erd\H{o}s-Renyi }
\newcommand{\uvec}{\mathbf{u}}

\begin{document}
\begin{frontmatter}

\title{A comparative study of social network models: network evolution models and nodal attribute models}

\author[BECS]{Riitta Toivonen\corref{cor1}}
\cortext[cor1]{Corresponding author}
\ead{Riitta.Toivonen@tkk.fi}
\author[BECS]{Lauri Kovanen}
\author[BECS]{Mikko Kivel\"{a}}
\author[OX,SAID,BECS]{Jukka-Pekka Onnela}
\author[BECS]{Jari Saram\"{a}ki}
\author[BECS]{Kimmo Kaski}
\address[BECS]{Department of Biomedical Engineering and Computational
  Science (BECS), Helsinki University of Technology, P.O. Box 9203,
  FIN-02015 HUT, Finland}
\address[OX]{Physics Department, Clarendon Laboratory, Oxford University, Oxford OX1 3PU, United Kingdom}
\address[SAID]{Sa\"id Business School, Oxford University, Oxford OX1
  1HP, United Kingdom}

\begin{abstract}
This paper reviews, classifies and compares recent models for social
networks that have mainly been published within the physics-oriented
complex networks literature. The models fall into two categories:
those in which the addition of new links is dependent on the
(typically local) network structure (\emph{network evolution models},
NEMs), and those in which links are generated based only on nodal
attributes (\emph{nodal attribute models}, NAMs). An exponential
random graph model (ERGM) with structural dependencies is included for
comparison.  We fit models from each of these categories to two
empirical acquaintance networks with respect to basic network
properties. We compare higher order structures in the resulting
networks with those in the data, with the aim of determining which
models produce the most realistic network structure with respect to
degree distributions, assortativity, clustering spectra, geodesic path
distributions, and community structure (subgroups with dense internal connections). We
find that the nodal attribute models successfully produce assortative
networks and very clear community structure. However, they generate
unrealistic clustering spectra and peaked degree distributions that do
not match empirical data on large social networks. On the other hand,
many of the network evolution models produce degree distributions and
clustering spectra that agree more closely with data. They also
generate assortative networks and community structure, although often
not to the same extent as in the data. The ERGM model turns out to
produce the weakest community structure. 
\end{abstract}

\end{frontmatter}

\section{\label{sec:intro}Introduction}

Modeling social networks serves at least two purposes. Firstly, it
helps us understand how social networks form and evolve. Secondly, in
studying network-dependent social processes by simulation, such as
diffusion or retrieval of information, successful network models can
be used to specify the structure of interaction. A large variety of
models have been presented in the physics-oriented complex networks
literature in recent years, to explore how local mechanisms of
network formation produce global network structure. In this paper we
review, classify and compare such models.

The models are classified into two main categories: those in which the
addition of new links is dependent on the local network structure
(\emph{network evolution models}, NEMs), and those in which the
probability of each link existing depends only on nodal attributes
(\emph{nodal attribute models}, NAMs).  NEMs can be further subdivided
into \emph{growing models}, in which nodes and links are added until
the network contains the desired number $N$ of nodes, and
\emph{dynamical models}, in which the steps for adding and removing
ties on a fixed set of nodes are repeated until the structure of the
network no longer statistically changes. For completeness, we include
in our comparative study two models from the tradition of exponential
random graph models (ERGMs). One of them is based solely on nodal
attributes, and the other incorporates structural dependencies. All of
these models produce undirected networks without multiple links or
self-links, and all networks are treated as unweighted, i.e. tie
strengths are not taken into account. We note that some of the models
were designed with a particular property in mind, such as a high
average clustering coefficient, but we will assess their ability to
reproduce several of the typical features of social networks. In
addition to comparing the distributions of degree and geodesic path
lengths and clustering spectra, we assess the presence or absence of
communities, which in the complex networks literature are 
typically defined as groups of nodes that are more densely connected
to nodes in the same community than to nodes in other communities~\cite{SantoReview}.

This paper is structured as follows. In Sections \ref{sec:NEMs} to
\ref{sec:ERGMs}, we define the categories of network evolution models
and nodal attribute models, and briefly review exponential random
graph models. Section \ref{sec:differences} discusses differences
between the philosophies behind NEMs and ERGMs. We fit models from
each of these categories to two empirical acquaintance networks with
respect to basic network statistics. The fitting procedure is
discussed in Section~\ref{sec:fitting} and Appendix~\ref{sec:appendix_fitting}. In
Section~\ref{sec:results}, we compare higher order structures in the
resulting networks with those in the
data. Section~\ref{sec:discussion} summarizes our results.

%---------------------------------------------------
\subsection{\label{sec:NEMs}Network evolution models (NEMs)}
%---------------------------------------------------

Let us first present a class of network models that focuses on network
evolution mechanisms. These models test hypotheses that specific
network evolution mechanisms lead to specific network structure. We
call these {\em network evolution models} (NEMs), and define them via
three properties as follows:
\begin{enumerate}[1)] 
  \item A single network realization $G$ is produced by an iterative
  process that always starts from an initial network configuration
  $G(t_0)$ specified in the NEM. Dynamical models often begin with an
  empty network, and growing models start with a small seed
  network\footnote{The seed network does not always need to be
  exactly specified, as long as it meets the given general criterion
  (such as being small compared to the network that will be
  generated), as it typically has a negligible effect on the
  resulting network.}.
  \item The specifications of the NEM include an explicitly defined
  set of stochastic rules by which the network structure evolves in
  time. These rules concern selecting a
  subset of nodes and links at each time step, and adding and deleting
  nodes and links within this subset. The rules typically correspond
  to abstracted mechanisms of social tie formation such as
  triadic closure~\citep{Granovetter}, i.e. tie formation based on
  the tendency of two friends of an individual to become acquainted.
  The rules always depend on network structure and they can sometimes
  also incorporate nodal attributes. The rules determine the possible
  transitions from one network $G(t_{k-1})$ to the next
  $G(t_k)$ during the iterative process that will produce one network
  realization $G=G(t_{end})$. 
  \item The NEM includes a stopping criterion:
  \begin{enumerate}[a)]
    \item For a {\em growing} NEM, the algorithm finishes when the
    network has reached a predetermined size. The typical assumption is
    that relevant statistical properties of the network remain
    invariant once the network is large enough.
    \item For a {\em dynamical} NEM, the algorithm finishes when
    selected network statistics no longer vary\footnote{While the
    stopping criterion for a growing NEM is exact, requirement 3b) is
    a heuristic criterion that assumes that the algorithm will reach a
    stage at which the selected statistical properties of the networks
    $G'(t)$ stabilize.  Although we cannot know with absolute
    certainty whether stationary distributions have been reached, we
    can be relatively confident of it if monitored properties remain
    constant and their distributions appear stable for a large number
    of time steps. 
}.
 \end{enumerate}
\end{enumerate}
  
A growing model can be motivated as a model for social networks in
several contexts. For example, on social networking sites people
rarely remove links, and new users keep joining the network.
Similarly, in a co-authorship network~\cite{NewmanCollaboration2001}
derived from publication records, existing links remain while new
links form. We point out that the growing models do not intend to
simulate the evolution of a social network ab initio.  However, the
mechanisms are selected to imitate the way people might join an
already established social network. 

The NEMs in our comparative study include only network structure
based evolution rules (that may depend on topology and tie strengths),
although nodal attribute based rules are also possible. Models in
which link generation is based {\em solely} on (fixed) nodal
attributes belong to the category of \emph{nodal attribute models}
(NAMs) discussed below.

% ______________________________________________________________
\subsection{Nodal attribute models (NAMs)}
\label{sec:NAMs}
% ______________________________________________________________

We adopt the term \emph{nodal attribute models} (NAMs)
for network models in which the probability of edge $e_{ij}$ between
nodes $i$ and $j$ being present is explicitly stated as a function of
the attributes of the nodes $i$ and $j$ only, and the evolutionary
aspect is absent. NAMs are often based on the concept of
\emph{homophily}~\citep{homophily}, the tendency for like to interact
with like, which is known to structure network ties of various types,
including friendship, work, marriage, information transfer, and other
forms of relationship. Such models have also been described by the
term {\em spatial models}~\citep{BPDA,WPR}, referring to that the fact
that the attributes of each node determine its 'location' in a social
or geographical space.

\subsection{Exponential random graph models (ERGMs)}
\label{sec:ERGMs}
{\em Exponential random graph models}
(ERGMs)~\citep{MarkovGraphs,Frank1991,WassermanPattison1996,ERGMintro,newSpecifications,ERGMRecentDevelopments},
also called $p^\ast$ models, are used to test to what extent nodal
attributes (exogenous factors) and local structural dependencies
(endogenous factors) explain the observed global structure. For
example, \citet{GoodreauERGMFittedToSchoolData} used ERGMs to infer
that much of the global structure (measured in terms of the
distributions of degree, edgewise shared partners and geodesic paths)
observed in a friendship network could be captured by nodal attributes
and patterns of shared partners and $k$-triangles, which are
relatively local structures.

Consider a random graph $\mathbf{X}$ consisting of $N$ nodes, in which
a possible tie between two nodes $i$ and $j$ is represented by a
random variable $X_{ij}$, and denote the set of all such graphs by
$\mathcal{X}$.  Using this notation, ERGMs are defined by the
probability distribution of such graphs $\mathbf{X}$
\begin{equation}
P_{ \theta, \mathcal{X}}(X=x) = \frac{ \exp{
 \{ \theta^t \, \uvec(x)  \} 
}  }{ c( \theta,\mathcal{X})},
\label{eq:ERGM}
\end{equation} 
where $\theta$ is the vector of model parameters, $\uvec(x)$ is a
vector of network statistics based on the network realization $x$, and
the denominator $c( \theta,\mathcal{X})$ is a {\em normalization}
function that ensures that the distribution sums up to one. The
selected statistics $\uvec(x)$ specify a particular ERGM
model. Typically, the parameters $\theta$ of an ERGM model are
determined using a maximum likelihood (ML) estimate, obtained by
Markov Chain Monte Carlo (MCMC)
sampling~\citep{MCMCML1992,Snijders2002}.  MCMC sampling heuristics
are also used to draw network realizations from the distribution
$P_{\theta, \mathcal{X}}$.  Several software packages are designed for
fitting and simulating ERGMs (including pnet, SIENA, and statnet,
discussed by~\citet{ERGMRecentDevelopments}).

\subsection{Differences between NEMs and ERGMs}
\label{sec:differences}
An important difference between network evolution models and
exponential random graph models is that a NEM is determined by the
rules of network evolution, whereas ERGMs do not explicitly address
network evolution processes. The particular update steps employed in
the iterative MCMC procedure for drawing samples are not explicitly
specified in ERGMs, which are defined by the probability distribution
$P_{\theta, \mathcal{X}}$, although MCMC methods can also be used to
model the evolution of social
networks~\citep{SnijdersActorOriented,SnijdersNetworkDynamics}.
A class of probability models that includes network evolution is the
stochastic actor-oriented models for network change proposed by
\citet{SnijdersActorOriented}, which are continuous-time
Markov chain models that are implemented as simulation models.
Another difference is that unlike ERGMs, NEMs explicitly specify an 
initial configuration from which the iteration is started, as well as 
a stopping criterion. However, NEMs are typically not sensitive to the
initial configuration. 

One of the known problems with ERGMs is that the distributions of
their sufficient statistics may be multimodal~\citep{Snijders2002}.
This has been of particular concern with respect to ERGMs that include
statistics related to transitivity, which is a highly relevant feature
in modeling social networks. The first stochastic model to express
transitivity, the {\em Markov graph}~\citep{MarkovGraphs}, employed a
simple triangle count term that is known to cause problems of model
degeneracy~\citep{Jonasson1999}, and to lead to instability in
simulation of large networks with Markov Chain Monte Carlo (MCMC)
methods~\citep{Snijders2002,HandcockDegeneracy,GoodreauERGMFittedToSchoolData}.
This problem seems to have been largely overcome with a recently
proposed term related to triangles, the {\em geometrically weighted
edgewise shared partners} statistic (GWESP)
\citep{newSpecifications,HunterGoodnessOfFit,ERGMRecentDevelopments}.
We include in our comparison an ERGM that includes the GWESP term. It
turns out that we encounter instability even with this model. In
fitting this model to our data, in the optimal parameter region
a very small modification of the model parameters produces a large
difference in the resulting network structure. This is discussed in
Section~\ref{sec:fitting} and Appendix~\ref{sec:appendix_fitting}.

In contrast, transitivity is easy to incorporate in NEMs. Problems of
multimodality have not been observed with NEMs. Although we do not
always have theoretical certainty that the network evolution rules
could not lead to multimodal distributions of network statistics, in
practice the models with given parameters seem to consistently produce
network realizations with similar statistics.

The NEMs and ERGMs lend themselves to testing different kinds of
hypotheses about networks. ERGMs can be employed to test to what
extent nodal attributes and local structural correlations explain the
global structure. Although both NEMs and ERGMs can easily incorporate
nodal attributes, they have rarely been included in NEMs. The NEMs
proposed so far have been of a fairly generic nature, whereas the ERGM
approach often aims to make inferences based on specific empirical
data, often including nodal attributes. On the other hand, NEMs can be
employed for testing hypotheses about network evolution, which ERGMs
do not explicitly address. For example, a NEM can be used to test
whether a combination of tie-strength-dependent triadic closure and
global connections can produce a clearly clustered
structure~\citep{KOSKK}. Although ERGMs can also be interpreted as
addressing endogenous (network structure based) selection processes
via structural dependencies, the mechanisms by which new ties are
created based on the existing network structure are made explicit only
in NEMs.

For the dynamical NEMs treated in this paper, it is easy to generate
(and estimate parameters for) networks of $10\,000$ nodes or more.
The growing models can easily produce networks with millions of nodes.
Based on our hands-on experience using state-of-the-art ERGM
software (statnet,~\citet{statnetProject,statnetArticle}), it seems
that generating a realization from a NEM might typically have much
lower computational cost than drawing a sample from an ERGM with
structural dependencies.  In generating network realizations from an
ERGM, we used as a guideline that the number of MCMC steps,
corresponding to the number of proposals for changes in the link
configuration, should be large enough such that the presence or
absence of a link between each dyad is likely to be changed several
times. With this approach, the number of MCMC steps should be
proportional to the number of dyads, implying that the complexity is
at least on the order of $O(N^2)$.  This is already a much larger
burden than the $O(N)$ complexity of NEMs based on local operations in
the neighborhood of a selected node. Our assumption of the
computational demands of ERGMs is supported by the fact that networks
that have thus far been studied with ERGMs have consisted typically of
at most a couple of thousands of nodes~\citep{GoodreauERGMFittedToSchoolData}.
\begin{figure}[htb]
\centering
\includegraphics[width=1.0\linewidth]{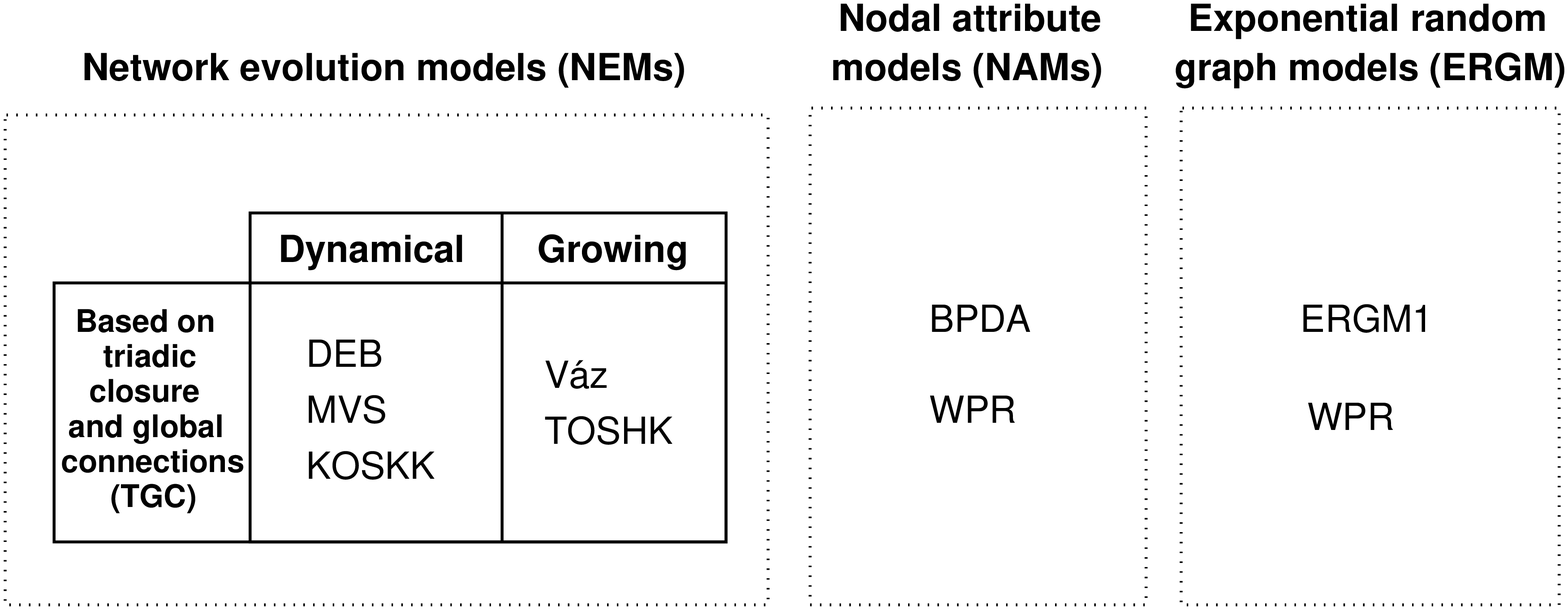}
\caption{Categories of social network models.  Within the category of
NEMs, we focus on models based on \tfc (\TFCn). Model labels
correspond to models discussed in Section~\ref{sec:models}. }
\label{fig:categories}
\end{figure}

\section{\label{sec:models}Description of the models}

Many complex networks models study the question of whether structures
observed in social networks could be explained by the
network-dependent interactions of nodes, without reference to
intrinsic properties of nodes. Such models are based on assumptions
about the local mechanisms of tie formation, such as people meeting
friends of friends, and thus forming connections with their network
neighbors (triadic closure~\citep{Granovetter}). An additional
mechanism to produce 'global' connections beyond the local
neighborhood is typically included to account for short average
geodesic path lengths~\citep{Milgram67}. Such connections may arise
from encounters at common hobbies, places of work, etc. In models that
do not consider nodal attributes, contacts between any dyads in the
network are considered equally likely. These two mechanisms, \tfc
(\TFCn), form the basis of all the NEMs we study in this work.

Tables \ref{mechanisms_dynamical}, \ref{mechanisms_growing} and
\ref{mechanisms_NAM} contain more detailed descriptions of the models
and their parameters, with fixed parameters given in
parentheses. Values of the fixed parameters were selected according to
the original authors' choices wherever possible.  We label the models
using author initials.

\begin{figure}[htb]
\centering
\includegraphics[width=0.7\linewidth]{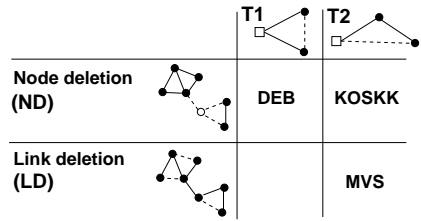}
\caption{The dynamical network evolution models DEB, KOSKK, and MVS, 
 classified according to the mechanisms for triadic closure and link
 deletion employed in them.}
\label{fig:fourfield_models}
\end{figure}

\paragraph{\label{sec:dynamical}Dynamical network evolution models}
We will first look at three dynamical models that combine \tfc (\TFCn)
for creating new links. These were proposed by \citet{DEB} (DEB),
\citet{MVS} (MVS), and \citet{KOSKK} (KOSKK). The different ways of
implementing triadic closure and deletion of links in each of these
models are highlighted in Fig.~\ref{fig:fourfield_models}.  In triadic
closure mechanism T1, a node is introduced to another node by their
common neighbor. In mechanism T2, new contacts are made through search
via friends: A node links to a neighbor of one of its neighbors.
Dynamical models in which new links are continuously added must also
include a mechanism for removing links, to avoid ending up with a
fully connected network.  In {\em node deletion} (ND), all links of a
node are deleted. This emulates a node 'leaving' and a newcomer
joining the network. In {\em link deletion} (LD), each link has a
given probability of being deleted at each time step.

The DEB model is the simplest of the three, with only two parameters,
network size $N$ and the probability $p$ of deleting a node. The MVS
and KOSKK models both use triadic closure mechanism T2, a two-step
search in the neighborhood of a node, but the KOSKK model takes
interaction strength into account. In KOSKK, new links are created
preferably through strong ties, and every interaction further
strengthens them. This mechanism is able to produce clear community
structure~\citep{KOSKK}, confirmed by our analysis in
Section~\ref{sec:results}. The three models also differ in whether a
new node can remain isolated for several time steps (as in the MVS
model) or will immediately link to another node (as in KOSKK), and in
whether there is a limit on the number of random connections each node
can make (as in DEB).  Because of such differences, it is difficult to
isolate the effects of the choices of T1 versus T2 and ND versus
LD. Therefore, in Section~\ref{sec:fourfield} we will combine the four
mechanisms using the DEB model as a basis. 

\citet{MVS} did not mention which value they used in the MVS model for
the probability $\lambda$ of deleting a link at each time step.  We
fixed $\lambda=0.001$ in our simulations, giving each tie an average
'lifetime' of $1000$ time steps.  When generating network
realizations, the dynamical models MVS, DEB, and KOSKK are iterated
until monitored distributions appear to become stationary. Sometimes
the authors do not state which particular criterion they used. For the
MVS and DEB models, we determined how many iterations (the steps
described in Table~\ref{mechanisms_dynamical}) it takes until average
degree stabilizes and its distribution appears stationary. When
generating networks, we used a number of iterations above this
limit. For the KOSKK model, we used a number of iterations determined
by the authors to be sufficient for the distributions of degree and
several other network properties to appear stationary ($2.5 \times
10^4 \times N$, where $N$ is network size, resulting in $2\times 10^8$
and $2.8 \times 10^7$ for fitting to our data sets of sizes $8003$ and
$1133$ presented in Section~\ref{sec:lastfm}.
\newcommand{\tablewidth}{0.99\linewidth}
\newcommand{\modelwidth}{0.12\linewidth}
\newcommand{\mechanismwidth}{0.99\linewidth}
\newcommand{\paramswidth}{0.13\linewidth}
\newcommand{\captionwidth}{0.97\linewidth}
\newcommand{\tablewidthSpatial}{0.98\linewidth}

\begin{table*}[htb]
\smallskip  \smallskip  \smallskip  
\caption{{\em Category: Dynamical network evolution models (dynamical NEMs)}. Three
models based on \tfcn. \smallskip  \smallskip }
\small
\begin{tabularx}{\tablewidth}{l X}  
 \hline
{\em Parameters} & {\em Mechanisms}.  Number of nodes $N$
  fixed; repeat steps for I) adding ties and II) deleting ties until stationary
  distributions are reached   \smallskip \\
\hline 
\multicolumn{2}{l}{ \textbf{DEB}  \citep{DEB}}  \\   
 \parbox{\paramswidth}{ 2 free \\ $N$,  $p$ } &
\parbox{\mechanismwidth}{
\smallskip
 I) Select a node $i$ randomly, and \\
\phantom{x} \parbox{\mechanismwidth}{a) if $i$ has fewer than two ties, introduce it to a random
node} \\
\phantom{x} \parbox{\mechanismwidth}{b) otherwise pick \emph{two neighbors} of $i$ and introduce them
if they are not already acquainted.} \\
II) Select a \emph{random node} and with prob.~$p$ \emph{remove all of its ties}. \smallskip
\smallskip
} \\
\hline   
\multicolumn{2}{l}{ \textbf{MVS}  \citep{MVS} }  \\ 
 \parbox{\paramswidth}{3 free \\ $N$,
$\xi$, $\eta$ \\ ($\lambda\!=\!0.001$) } &
 \parbox{\mechanismwidth}{
\smallskip
I) Select a node $i$ randomly, and \\
 \phantom{x} \parbox{\mechanismwidth}{ a) connect $i$ to another random
 node with probability~$\eta$.}\\
 \phantom{x} \parbox{\mechanismwidth}{ b) select a \emph{friend's
 friend} of $i$ (by uniformly random search) with probability~$\xi$ 
and introduce $i$ to it if not already acquainted.\smallskip } \\
II) Select a \emph{random tie} and delete it with probability~$\lambda$.  \smallskip
\smallskip} \\  
\hline 
\multicolumn{2}{l}{ \textbf{KOSKK}  \citep{KOSKK}}  \\ 
\parbox{\paramswidth}{ $3$ free \\ $N$,  $p_{\Delta}$, $p_r$
\\ 
($w_0 =1$, \\ 
\mbox{$p_d=0.001$}, \\
$\delta=0.5$)
} &
\parbox{\mechanismwidth}{
\smallskip
I) Select a node $i$ randomly, and \\
 \phantom{x} \parbox{\mechanismwidth}{ a) select a \emph{friend's
 friend} $k$ (by \emph{weighted
  search}) and introduce it to $i$ with prob.~$p_{\Delta}$  (with
 initial tie strength  $w_0$) if not
  already acquainted. \emph{Increase tie strengths} by $\delta$ along the search
  path, as well as on the link $l_{ik}$ if it was already present. \smallskip }
\phantom{x}  \parbox{\mechanismwidth}{ b) additionally, with
 prob.~$p_r$ (or with prob. $1$ if $i$ has no connections), connect
 $i$ to a random node $j$ (with tie strength $w_0$). }\\ 
II) Select a
 \emph{random node} and with prob.~$p_{d}$ \emph{remove all of its
 ties}. \smallskip
\smallskip} \\   
\hline 
\multicolumn{2}{l}{ 
\parbox{\captionwidth}{ \smallskip \smallskip Nodes represent individuals
 and links represent ties between them. Parameters whose values were fixed
 according to the original authors' choices are shown in parentheses.}
 }  \\
\end{tabularx}
\label{mechanisms_dynamical}
\end{table*}

\paragraph{Growing network evolution models}
We include two growing models, proposed by \citet{VAZ} (\vazn) and
\citet{TOSHK} (TOSHK). They are described in detail in
Table~\ref{mechanisms_growing}. These are to our knowledge the only
growing models specifically proposed for social acquaintance networks.
The motivation behind the \vaz model is to produce a high level of
clustering and a power law degree distribution. The TOSHK model also
aims at a broad degree distribution and a high clustering coefficient,
but also sets out to reproduce other features observed in social
networks, such as community structure.

In TOSHK, each new node links to one or more 'initial contacts', which
in turn introduce the newcomer to some of their neighbors. In \vazn, a
newcomer node first links to a random node $i$, creating {\it
potential edges} (V{\'a}zquez's term) between itself and the neighbors
of $i$. These ties may be realized later, generating triangles in the
network. In both models, triangles are only generated between the
newcomer and the neighbors of its initial contact, and further
processes of introduction are ignored. As with all the models, we keep
to the authors' choices presented in the original paper. Accordingly,
in the TOSHK model, we allow a newcomer to link to at most two initial
contacts (see Table~\ref{mechanisms_growing}), and pick the number of
secondary contacts from the uniform distribution $U[0,k]$, although
this clearly limits the adaptability of the model.

\begin{table*}[htb]
\caption{{\em Category: Growing network evolution models (growing NEMs)}. Two models based
  on \tfcn. \smallskip  \smallskip }
\small
\begin{tabularx}{\tablewidth}{l X}\hline 
{\em Parameters} & {\em Mechanism}. Repeat steps for I) adding nodes
  and ties II) adding ties only until network contains $N$ nodes.  \smallskip \\
\hline
\multicolumn{2}{l}{ \textbf{TOSHK}  \citep{TOSHK} }  \\
  \parbox{\paramswidth}{ 3 free \\  $N$, $p$, $k$\\ (simplified) }  &
 \parbox{\mechanismwidth}{\smallskip \smallskip 
I) Add a new node $i$ to the network, connecting it to one
random initial contact with probability $p$, or two with probability $1-p$.\\
II) For each random initial contact $j$, draw a number $m_{sec}$
of secondary connections from the distribution $U[0,k]$ and connect $i$ to $m_{sec}$
neighbors of $j$ if available.
\smallskip \smallskip }\\  
\hline
\multicolumn{2}{l}{ \textbf{V\'az}   \citep{VAZ} }  \\ 
\parbox{\paramswidth}{2 free \\  $N$, $u$ }  &
 \parbox{\mechanismwidth}{\smallskip I) With probability $1-u$, add a
new node to the network, connecting it to a random node
$i$. Potential edges are created between the newcomer $n$ and the
neighbors $j$ of $i$ (a potential edge means that $n$ and $j$ have a
common neighbor, $i$, but no direct link between them).\\ II) With
probability $u$, convert one of such potential edges generated on any
previous time step to an edge. 
Potential edges generated by converting an edge are ignored.
\smallskip \smallskip }\\  
\hline
\end{tabularx}
\label{mechanisms_growing}
\end{table*}

\paragraph{Nodal attribute models}
 We study two nodal attribute models that differ in the dependence of
link probability on distance and in the employed distance
measure. These models, proposed by \citet{BPDA} (BPDA) and \citet{WPR}
(WPR), are described in Table~\ref{mechanisms_NAM}.  The authors
mention that a social space of any dimension could be used, but study
the cases of 1D and 2D, respectively. We keep to their choices. 

\begin{table*}[htb]
\smallskip  \smallskip  \smallskip
\caption{ {\em Category:  Nodal attribute models (NAMs)}.  \smallskip  \smallskip}
\small
\begin{tabularx}{\tablewidthSpatial}{l X}\hline
{\em Parameters} & {\em Mechanism}  \smallskip   \\
\hline 
\multicolumn{2}{l}{ \textbf{BPDA}  \citep{BPDA}}   \\ 
\parbox{\paramswidth}{3 free \\ $N$, $\alpha$, $b$  } &
 \parbox{\mechanismwidth}{\smallskip 
Distribute $N$  nodes with uniform probability in a (1-dimensional) social space (a
segment of length $h_{max}$).
Link nodes with prob.~$p= 1 / \left( 1+ \left( d/b \right)^{\alpha}
\right) $,
where $d$ is their distance in the social space. ($h_{max}$ can be
absorbed within $b$). If treated many-dimensionally, similarity along
one of the social dimensions is sufficient for the nodes to be seen as
similar.
  \smallskip \smallskip}\\  
\hline
\multicolumn{2}{l}{ \textbf{WPR}  \citep{WPR}}   \\   
\parbox{\paramswidth}{ 4 free \\ \mbox{$N$, $H$, $p$, $p_b$} 
}  &
 \parbox{\mechanismwidth}{\smallskip Distribute $N$ nodes according to
a homogeneous Poisson point process in a (2-dimensional) social space
of unit area. Create a link between each node pair separated by
distance $d$ with probability~$p+p_b$ if $d<H$, and with
probability~$p-p_{\Delta}$ if $d>H$ (where $p_{\Delta}(p,p_b,H)$ is
such that the total fraction $p$ of all possible links is generated).
\smallskip \smallskip}\\    
\hline 
\end{tabularx}
  \label{mechanisms_NAM}
\end{table*}

\paragraph{ERGM with structural dependencies} 
As our data does not contain nodal attributes, we can only include
structural terms in the exponential random graph model labeled ERGM1
(Table~\ref{mechanisms_ERGM}). The term edge count is an obvious 
choice to include, in order to match average degree. We must also
include a term related to triads, considering the prevalence of
transitivity social networks. We employ the geometrically weighted
edgewise shared partner statistic (GWESP), proposed by
\citet{newSpecifications} and formulated by
\citet{HunterGoodnessOfFit} as
\begin{equation} \label{eq:v}
v(x;\tau)= e^{\tau} \sum_{i=1}^{n-2} \{ 1 - (1-e^{\tau})^i \} EP_i(x),
\end{equation}
where the {\em edgewise shared partners statistic} $EP_i(x)$ indicates
the number of unordered pairs $\{ j, k\}$ such that $x_{jk}=1$ and $j$
and $k$ have exactly $i$ common neighbors~\citep{curvedERGM}.  The
simple triangle count term employed in Markov random graphs is known
to cause problems of multimodality, and we are not aware of other
triangle-related terms that would have been employed in ERGMs.
Because we would also like to match the degree distribution to the
data, we include the geometrically weighted degree term (GWD)
\citep{newSpecifications,HunterGoodnessOfFit}\footnote{%
\citet{GoodreauERGMFittedToSchoolData} observed that the model
edges+covariates+GWESP explains much of the observed data (an
adolescent friendship network with 1681 actors) and that no
improvement is achieved by including the terms geometrically weighted
degree (GWD) or geometrically weighted dyadwise shared partners
statistic (GWDSP). Based on this, it seems that the terms GWD and
GWDSP might not bring additional value to a model that already
includes the GWESP term.  However, the conclusions drawn by
\citet{GoodreauERGMFittedToSchoolData} might not be transferable to
our case because our data is different; for example, we do not have
nodal attribute data. }
\begin{equation} \label{eq:u} 
u(x;\tau)= e^{\tau} \sum_{i=1}^{n-2} \{ 1 - (1-e^{\tau})^i\} D_i(x),
\end{equation}
where $D_i$ indicates the number of nodes with degree $i$. We fix the
parameter $\tau=0.25$ as in~\citep{GoodreauERGMFittedToSchoolData}.
We generate network realizations using the statnet
software~\citep{statnetArticle}. MCMC iterations are started from an
\er (Bernoulli) network with average degree matching the target. We
draw 5 realizations from each MCMC chain at intervals of $10^7$, using
a burn-in of $5 \times 10^7$ time steps. Model parameters are
optimized consistently for all models with the procedure described in
Section~\ref{sec:fitting} and Appendix~\ref{sec:appendix_fitting}.

\begin{table*}[htb]
\caption{ {\em Category: exponential random graph models (ERGM) with
    structural dependencies}.  \smallskip \smallskip} \small
\begin{tabularx}{\tablewidthSpatial}{l X}\hline
{\em Parameters} & {\em Definition}  \smallskip   \\
\hline 
\multicolumn{2}{l}{ \textbf{ERGM1}  \citep{newSpecifications}}   \\ 
\parbox{\paramswidth}{4 free \\ $N$, $\theta_{L}$,\\ $\theta_{GWESP}$, $\theta_{GWD}$ \\  ($\tau=0.25$) } & 
 \parbox{\mechanismwidth}{\smallskip 
The model is defined
with three terms: edge count $L$,  geometrically weighted edge-wise shared partners (GWESP) $v(x;\tau)$ (Eq.~\ref{eq:v}),
% $v(x;\tau)= e^{\tau} \sum_{i=1}^{n-2} \{ 1 - (1-e^{\tau})^i \} EP_i(x)$, 
and geometrically weighted degree (GWD) $u(x;\tau)$ (Eq.~\ref{eq:u}),
% $u(x;\tau)= e^{\tau} \sum_{i=1}^{n-2} \{ 1 - (1-e^{\tau})^i \} D_i(x)$,
as the probability distribution 
\[
 P_{ \theta, \mathcal{X}}(X=x) = \frac{ \exp{
 \{
 \theta_L L + \theta_{GWESP} v(x) + \theta_{GWD} u(x)
 \}
}  }{ c( \theta,\mathcal{X})} 
\]
%\mbox{$P_{\theta, \mathcal{X}}(X=x) = \frac{ \exp{ \{ \theta \, u(x)  \} }  }{ c(\theta,\mathcal{X})}$ }
% displayed in Eq.~\ref{eq:ERGM},
% with the vector $\mathbf{u}(x)$ containing 
  \smallskip \smallskip}\\  
\hline
\multicolumn{2}{l}{ 
\parbox{\captionwidth}{  \smallskip \smallskip } \smallskip \smallskip \smallskip}  \\
\end{tabularx}
  \label{mechanisms_ERGM}
\end{table*}

% $N=1160$, $\theta_L=-6.962$, $\theta_{GWESP}=2.4$, $\theta_{GWD}=0.225$

\section{\label{sec:fitting}Fitting the models}
In order to compare networks generated by different models, it is
necessary to unify some of their properties. To this end, we fit the
models to two real-world data sets with respect to as many of the most
relevant network features as the model parameters allow. Our fitting
method consists of simulating network realizations with different
model parameters, and finding the parameter values that produce the
best match to selected statistics.

\subsection{Targeted features for fitting} 
The most important properties that we wish to align between the models
and the data are the number of nodes and links. Because both of our
data sets are connected components of a larger network, we focus on
the properties of the largest connected component of the generated
networks. Our first two fitting targets are largest connected
component size $N_{LC}$ and the average number of links per node, or
average degree \kaven, within the largest component. They are already
sufficient for fitting the DEB and \vaz models, which have only two
parameters.  A natural choice for the next target is some measure
related to triangles, because they are highly prevalent in social
networks. We will use the average clustering coefficient \cave (please
see Appendix~\ref{sec:characteristics} for the definition), which is a
well-established characterization of local triangle density in the
complex networks literature. All of the network evolution models in
this study had as one of their aims obtaining a high clustering
coefficient. These three features are sufficient for fitting the rest
of the models except WPR, if we fix some of the parameters according
to the original authors' choices (please see
Table~\ref{mechanisms_dynamical}).

If matching $N_{LC}$, \kave and \cave is not enough to fix all
parameters of the model, we no longer have a straightforward
choice. We considered using the assortativity coefficient and geodesic
path lengths~(see \ref{sec:characteristics}). In the WPR model,
assortativity varies closely together with the average clustering
coefficient, so it could not be used as a fourth target
feature. Instead, we used the average geodesic path length. We also
attempted using the assortativity coefficient for fitting the KOSKK
model, allowing the weight increment parameter $\delta$ to vary, but
ran into a different problem: attempting high assortativity forced the
weight increment parameter to zero, thereby eliminating an important
feature of the weighted model and weakening the community structure.
Hence, we fixed $\delta=0.5$ in accordance with the authors' choice.

All of these measures - degree, high clustering, assortativity, and
geodesic path lengths - assess important properties of social
networks, which are likely to affect dynamics such as opinion
formation or spreading of
information~\citep{mobile1,rumor,socialDynamicsCastellano}. The
average properties can typically be tuned by varying parameter values,
but the general shapes of the distributions are likely to be
invariable.

\subsection{\label{sec:lastfm}The friendship network at www.last.fm and the email network}
We selected two social network data sets with slightly different
average properties, in order to get a better picture of the
adaptability of the models. They differ in average degree, average
clustering coefficient, and the assortativity coefficient, although
both display assortativity and high clustering. 

We collected a mutual friendship network of users of the web service
{\em last.fm}. At the web site \emph{www.last.fm}, people can share
their musical tastes and designate other users as their friends. We
used for this study only the friendship information, disregarding the
musical preferences. Because there are several hundred thousand users
on the site worldwide, we selected users in one country, Finland, to
obtain a smaller network with $8003$ individuals.  The country labels
were self-reported. This data set (henceforth called \LFFn) represents
the largest connected component of Finnish users at this
site. Individuals in the resulting network have on the average \kave
$=4.2$ friends, and a high clustering coefficient \cave $= 0.31$. The
network is highly assortative with $r=0.22$, indicating that friends
of those users who have many connections at the site are themselves
well connected (please see Appendix~\ref{sec:characteristics} for
definitions). After designating someone as a friend, there is no cost
to maintaining the tie, i.e. the link never expires. This means that
the data may overestimate the number of active friendships within the
last-fm web site. However, the degree distribution is not broader than
that observed in a network constructed from mobile phone
calls~\citep{mobile2}, in which each contact has a real cost in time
and money. Requiring ties to be reciprocated ensures that the users
have at least both acknowledged one another.

We also use a smaller acquaintance network collected 
by~\citet{emailRef}, based on emails between members of the University
Rovira i Virgili (Tarragona). In the derived network, two individuals
are connected if each sent at least one email to the other during the
study period, and bulk emails sent to more than $50$ recipients are
eliminated. Again, we use the largest connected component of the
network. It consists of $1133$ individuals, and it is a compact
network with average geodesic path length \lave$=3.6$, 
average degree \kave$=9.6$, fairly high average clustering coefficient
\cave$=0.22$, and fairly small assortativity $r=0.08$.

\begin{figure}[htb]
\centering  
\includegraphics[width=1.0\linewidth]{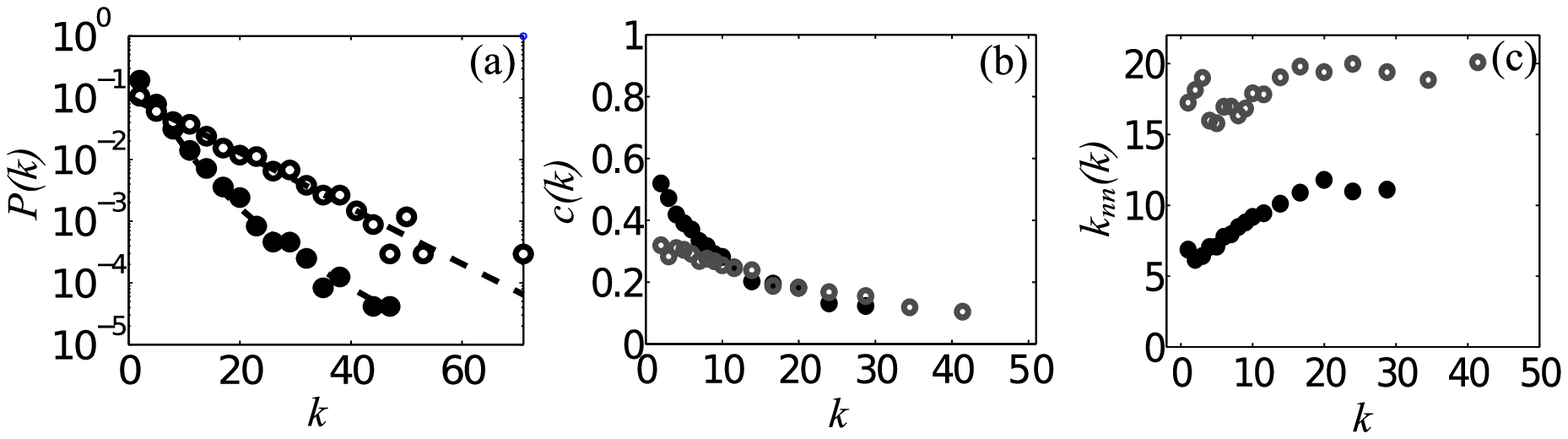}
\caption{Properties of the \LFF data set ({\Large $\bullet$}) and the
\email data ({\Large $\circ$}). a) degree distributions, with average
degrees \kave $= 4.2$ and $9.6$, respectively.  ~\citet{emailRef}
fitted to the \email data an exponential distribution
$p(k)=e^{-k/k^{\ast}}$ with $k^{\ast}=9.2$, which shows as a straight
line in a semilogarithmic plot.  The lognormal distribution fitted the
\LFF data best of the different distributions we tried (exponential,
Weibull, gamma, and lognormal), although not perfectly.  b) Clustering
$c(k)$ decreases with degree $k$ (average clustering \cave $=0.31$ and
$0.22$, respectively). c) Degree-degree correlations between nodes and
their neighbors ($k_{nn}$ signifies average nearest neighbor degree)
show that both networks are assortative (with $r=0.22$ and $r=0.08$,
respectively).}
\label{fig:data}
\end{figure}

Both of our empirical networks are unweighted, meaning that tie 
strengths are not specified. All of the models studied here apart from
KOSKK are unweighted as well. Averaged basic statistics of both data
sets are displayed in Table~\ref{table:lastfm_email}. The degree
distributions, clustering spectra and degree-degree correlations of
the \LFF and \email networks are shown in Fig.~\ref{fig:data}, and
more plots of their statistics are shown in Section~\ref{sec:results} in
connection with the fitted models.

Table~\ref{table:targeted_features} indicates which features were
targeted when optimizing the parameters of each model, and displays
the optimized parameters. Table~\ref{table:lastfm_email} displays
properties of the networks generated with these parameters. Due to the
stochastic nature of the models, two network realizations generated
with the same parameters are not likely to have exactly the same
average properties.  The plots and tables concerning the model
networks in this paper always contain values averaged over $100$
network realizations.

Fitting to a limited number of data sets does not allow full
assessment of the adaptability of the models.  However, the features
that we examine are similar in our two data sets as in other large
scale empirical social networks, such as those based on communication
via mobile phone~\citep{mobile2,mobile_DPLN} and Microsoft
Messenger~\citep{MSnetwork}. For example, these networks have skewed
degree distributions that imply the presence of high degree nodes,
high average clustering coefficients \caven, decreasing clustering
spectra $c(k)$, and positive degree-degree-correlations $r$. A
detailed description of the fitting procedure is included in Appendix~\ref{sec:appendix_fitting}.

\subsection{\label{sec:adaptability}Adaptability of the models}
Not surprisingly, for almost all models, average largest component
size \nlave and average degrees \kave could be fitted closely to both
data sets. For the models with only two free parameters (DEB, V\'az),
we had no control over other network features. These two-parameter
models turn out to have excessively high average clustering
coefficients for the moderate average degrees displayed in our two
data sets. For most of the other models, clustering could be tuned
rather closely. The TOSHK model, with its discrete parametrization of
the number of triangles formed, was not able to exactly match the
clustering values despite having three parameters.  

For the model ERGM1, we allowed the average degree to remain slightly
below the target in order to obtain correct clustering, because aiming
at both correct average degree and clustering led to an instable
region of model parameters.  We initially attempted using automated
optimization algorithms (such as snobfit~\citep{snobfit}) to fit the
ERGM1 model, but these failed due to the instability. Based on the
intuition of the model parameters obtained from the attempts at
fitting, we initially selected values that roughly produced the
desired \nlaven, \kaven, and \caven, and manually modified them for a
better fit. Starting from parameter values that generated networks in
which the clustering coefficient matched the \email data and the
average degree was only slightly too small, it turned out that a very
small increase in the parameter $\theta_L$ (done in order to increase
average degree) caused average degree to jump dramatically and the
clustering coefficient to plummet (see Fig.~\ref{fig:ERGM_transition}
in Appendix~\ref{sec:appendix_fitting}). Hence, we settled for a lower value of \kaven.

Average geodesic path lengths \lave were approximately correct for all
but the nodal attribute model treated in one dimension (BPDA),
although \lave was used for fitting only in the WPR model.  The
assortativity coefficient $r$ was not used for fitting any model,
although we attempted using it for fitting WPR and ERGM1. The ERGM1
model was only fitted to the \email data, because generating networks
of size $8000$ and fitting their parameters did not seem feasible for
the ERG model. 

\newcommand{\fittedParamsWidth}{10.0cm}
\begin{centering}
\begin{table}[ht]
\caption{Targeted network features, and the fitted model parameters leading to the
values closest to the \LFF and \email data sets. \smallskip }
\tiny
\begin{tabularx}{\tablewidth}{lX} 
\hline
 DEB &  matched to \nlaven,  \kave  \\
& \parbox{\fittedParamsWidth}{ \emph{lastfm}:  $N=8330,  p=0.203 $ \\  \emph{email}:   $\,N=1138,  p=0.064 $  }    \smallskip   \\ 
%
%\hline 
 MVS &  matched to \nlaven,  \kave, \cave   \\
 & \parbox{\fittedParamsWidth}{ \emph{lastfm}:  $N=9300,  \,  \xi=0.0022,  \,   \eta=  0.000368$ \\  \emph{email}:   $\,N=2270,  \,  \xi=0.0062,  \,    \eta=  0.000071$  }    \smallskip   \\ 
%
%\hline 
 KOSKK &  matched to \nlaven,  \kave, \cave \\
& \parbox{\fittedParamsWidth}{ \emph{lastfm}:  $N=8205,  p_{\Delta}=  0.0029,  \,  p_{r}=0.0008  $ \\  \emph{email}:   $\,N=1135,  p_{\Delta}=  0.0107,  \,  p_{r}=0.0039$  }    \smallskip   \\ 
%
%\hline 
TOSHK &  matched to $N$,  \kave, \cave \\
& \parbox{\fittedParamsWidth}{ \emph{lastfm}:  $N=8003, p=0.60, k=1 $ \\  \emph{email}:   $\,N=1133,\, p=0.06, k=3 $  }    \smallskip   \\ 
%
%\hline 
V\'az &  matched to $N$,  \kave \\
& \parbox{\fittedParamsWidth}{ \emph{lastfm}:  $N=8003, u=0.524 $ \\  \emph{email}:   $\,N=1133,\, u=0.793 $  }     \smallskip  \\ 
%
%\hline 
ERGM1 &  matched to \nlaven,  \kave, \cave  \\
& \parbox{\fittedParamsWidth}{ \emph{lastfm}: $-$ \\  \emph{email}:    $N=1160, \theta_{L}=-6.962$,
$\theta_{GWESP}=2.4$, \\$\theta_{GWD}=0.225$ }    \smallskip   \\ 
%
%\hline 
BPDA &  matched to \nlaven,  \kave, \cave \\
& \parbox{\fittedParamsWidth}{ \emph{lastfm}:  $N=8250, \alpha=1.915,\, b=1.51\cdot 10^{-4}$\\  \emph{email}:    $\,N=1133,\alpha=1.565,\,b=0.002032$  }   \smallskip  \\  
%
%\hline 
WPR &  matched to \nlaven,  \kave, \cave, \lave  \\
& \parbox{\fittedParamsWidth}{ \emph{lastfm}:  $N=8200, H=0.0108, p=0.000506, p_b=0.9994 $\\  \emph{email}:   $\,N=1133,\, H=0.040,\, p=0.008498,\, p_b=0.991 $ }    \smallskip  \\ 
\hline 
\multicolumn{2}{l}{ 
\parbox{\captionwidth}{ \smallskip \nlaven: average largest component size
(number of nodes),
\kaven: average degree,  \caven: average clustering coefficient,
\laven: average shortest path length. \kaven, \caven, and \lave were
calculated for the largest component of the network. } }  \\
\end{tabularx}
\label{table:targeted_features}
\end{table}
\end{centering}

\begin{table*}[ht]
\caption{Basic statistics of the \LFF and \email data sets and the
  models fitted to each.  \smallskip} \tiny
\begin{tabularx}{0.9\linewidth}{Xlllllll}\hline  \smallskip \smallskip
{ \it model / data } & \nl & $L$ & \kave & \cave & $r$  & \lave &
 $l_{max}$ \smallskip \smallskip \\ 
 Last-fm-fin						 & 8003 & 16824 & 4.20 & 0.31 & 0.22 & 7.4 & 24\\
\parbox{\modelwidth}{DEB                                                 }	 & 8009 $\pm $ 30 & 16858 $\pm $ 224 & 4.21 $\pm $ 0.05 & 0.38 $\pm $ 0.01 & 0.10 $\pm $ 0.01 & 7.0 $\pm $ 1.6 & 18.1 $\pm $ 1.4\\
\parbox{\modelwidth}{MVS                                                 }	 & 7989 $\pm $ 38 & 16816 $\pm $ 153 & 4.21 $\pm $ 0.03 & 0.30 $\pm $ 0.01 & 0.02 $\pm $ 0.01 & 7.8 $\pm $ 1.6 & 17.4 $\pm $ 1.0\\
\parbox{\modelwidth}{KOSKK                                               }	 & 8006 $\pm $ 20 & 16849 $\pm $ 207 & 4.21 $\pm $ 0.05 & 0.31 $\pm $ 0.01 & 0.05 $\pm $ 0.01 & 7.2 $\pm $ 1.5 & 16.3 $\pm $ 0.9\\
\parbox{\modelwidth}{TOSHK                                               }	 & 8003  & 16791 $\pm $ 93 & 4.20 $\pm $ 0.02 & 0.34 $\pm $ 0.01 & 0.14 $\pm $ 0.01 & 6.6 $\pm $ 1.3 & 13.8 $\pm $ 0.6\\
\parbox{\modelwidth}{Vàz                                                 }	 & 8003  & 16801 $\pm $ 171 & 4.20 $\pm $ 0.04 & 0.29 $\pm $ 0.01 & 0.27 $\pm $ 0.02 & 8.3 $\pm $ 2.6 & 22.6 $\pm $ 1.5\\
\parbox{\modelwidth}{BPDA                                                }	 & 8005 $\pm $ 31 & 16794 $\pm $ 141 & 4.20 $\pm $ 0.03 & 0.29 $\pm $ 0.01 & 0.30 $\pm $ 0.02 & 23.9 $\pm $ 9.3 & 60.1 $\pm $ 8.0\\
\parbox{\modelwidth}{WPR
 }	 & 8004 $\pm $ 19 & 16972 $\pm $ 150 & 4.24 $\pm $ 0.03 & 0.29
 $\pm $ 0.01 & 0.30 $\pm $ 0.02 & 8.1 $\pm $ 1.6 & 18.2 $\pm $ 1.1  \smallskip \smallskip  \\
 \hline \smallskip \smallskip 
{ \it model / data } & \nl & $L$ & \kave & \cave & $r$  & \lave &
 $l_{max}$  \smallskip \smallskip  \\ 
Email						 & 1133 & 5451 & 9.62 & 0.22 & 0.08 & 3.6 & 7\\
\parbox{\modelwidth}{DEB                                                 }	 & 1133 $\pm $ 3 & 5452 $\pm $ 249 & 9.62 $\pm $ 0.43 & 0.45 $\pm $ 0.01 & 0.06 $\pm $ 0.02 & 3.4 $\pm $ 0.9 & 7.7 $\pm $ 0.7\\
\parbox{\modelwidth}{MVS                                                 }	 & 1113 $\pm $ 1 & 5282 $\pm $ 77 & 9.48 $\pm $ 0.14 & 0.23 $\pm $ 0.01 & 0.05 $\pm $ 0.04 & 3.8 $\pm $ 1.1 & 9.6 $\pm $ 0.6\\
\parbox{\modelwidth}{KOSKK                                                }	 & 1134 $\pm $ 2 & 5425 $\pm $ 193 & 9.57 $\pm $ 0.34 & 0.22 $\pm $ 0.01 & 0.06 $\pm $ 0.02 & 3.5 $\pm $ 0.9 & 7.5 $\pm $ 0.6\\
\parbox{\modelwidth}{TOSHK                                                }	 & 1133  & 5453 $\pm $ 52 & 9.63 $\pm $ 0.09 & 0.29 $\pm $ 0.01 & 0.09 $\pm $ 0.02 & 3.4 $\pm $ 0.8 & 6.1 $\pm $ 0.3\\
\parbox{\modelwidth}{Vàz                                                  }	 & 1133  & 5453 $\pm $ 136 & 9.63 $\pm $ 0.24 & 0.42 $\pm $ 0.02 & 0.12 $\pm $ 0.03 & 4.6 $\pm $ 1.7 & 13.6 $\pm $ 1.4\\
\parbox{\modelwidth}{BPDA                                                }	 & 1133 $\pm $ 1 & 5477 $\pm $ 172 & 9.67 $\pm $ 0.30 & 0.22 $\pm $ 0.01 & 0.22 $\pm $ 0.02 & 4.4 $\pm $ 0.8 & 8.4 $\pm $ 0.5\\
\parbox{\modelwidth}{WPR                                                 }	 & 1133 $\pm $ 1 & 5448 $\pm $ 72 & 9.62 $\pm $ 0.13 & 0.21 $\pm $ 0.01 & 0.20 $\pm $ 0.03 & 3.6 $\pm $ 0.7 & 6.0 $\pm $ 0.2  \\
% add here ERGM1 model       model_ERGM/params_N1160_-6.962_2.4_0.225
\parbox{\modelwidth}{ERGM1 }	  & 1133  $\pm $ 8  & 4800 $\pm $ 460  & 8.47 $\pm $ 0.77 & 0.21   $\pm $  0.01      & 0.04   $\pm $   0.02   & 3.6 $\pm $  0.84  & 7.5 $\pm $ 0.83 \smallskip \smallskip \\ 
 \hline 
\multicolumn{8}{l}{ 
\parbox{\captionwidth}{ \small \smallskip All statistics are
  calculated for the largest component of each network. $N_{LC}$:
  Largest component size, $L$: number of links, \kaven: average
  degree, \caven: average clustering coefficient, $r$: assortativity
  coefficient, \laven: average geodesic path length, and $l_{max}$:
  longest geodesic path length.  The values are averaged over $100$
  realizations of each network model. The standard error of the
  averages is displayed whenever there was fluctuation in the values.
  \smallskip \smallskip } } \\
\end{tabularx} 
\label{table:lastfm_email} 
\end{table*}

\section{\label{sec:results}Comparison of higher order statistics}
Having fitted the models according to average values of particular
network characteristics, we address their degree distributions $P(k)$,
clustering spectra $c(k)$, and geodesic path length distributions
$P(l)$.  We also assess the community structure of the networks using
several measures. In Section~\ref{sec:fourfield} we combine and
compare the different mechanisms for triadic closure and link deletion
employed in the dynamical NEMs.  We use graphs to assess goodness of
fit as promoted by~\citet{HunterGoodnessOfFit}.

\subsection{\label{sec:results_degree}Degree distribution} 
\begin{figure}[htb]
\centering
\includegraphics[width=1.0\linewidth]{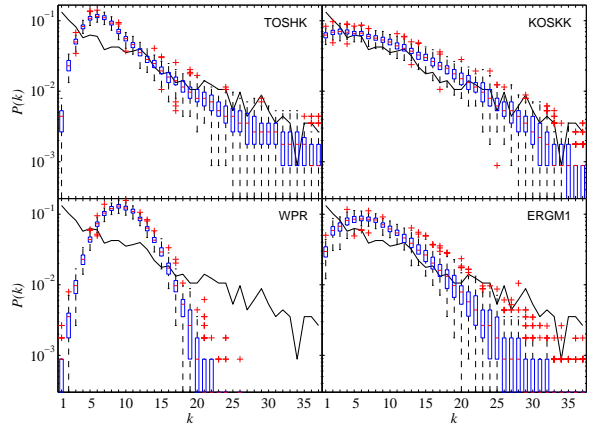}
\caption{Degree distributions $P(k)$ of the \email data (solid line)
 and in selected models fitted to it. The box plots display medians
 and first and third quartiles in $100$ network realizations.
 Whiskers extend from each end of the box to the most extreme values
 in the data within $1.5$ times the interquartile range from the ends
 of the box. Outliers are denoted by \textcolor{red}{$+$}. } 
\label{fig:degreeDistributions}
\end{figure}

Degree distributions are shown in Fig.~\ref{fig:degreeDistributions}
for the \email data and selected models.  The exact shapes of the
degree distributions produced by the models are not as important as
their markedly different probabilities for the presence of high degree
nodes (Fig.~\ref{fig:degreeDistributions}).  The nodal attribute
models, of which the \LFF fit of WPR is shown, produce skewed but
fast-decaying degree distributions that imply the absence of nodes
with very high degree.  These distributions are well fit with the
Poisson distribution\footnote{The homophily principle does not always
lead to a Poisson degree distribution. The shape of the degree
distribution depends on how the nodal attributes are
distributed. \citet{vip} used an exponentially distributed fitness
parameter as the basis for homophily, and obtained a flat degree
distribution P(k)=const. As they observe, this is
unrealistic. Combined with another mechanism, homophily can also lead
to a broader degree distribution \citep{vip}.}, as shown analytically
by \citet{BPDA} for the BPDA model.  The \vaz model produces a very
broad degree distribution (not shown) that was shown by \citet{VAZ} to
decay as power law, $P(k) \sim k^{- \gamma}$, which implies the
presence of a few nodes with extremely high degree.  The tails of the
degree distributions produced by the dynamical NEMs and the growing
TOSHK model as well as the ERGM1 model all appear to decay slower than
the Poisson distribution, but faster than power law. Of these, the
models TOSHK, KOSKK, and ERGM1 are displayed in
Fig.~\ref{fig:degreeDistributions}.

In our data sets, the degree distribution decays exponentially
(\emailn)~\citep{emailRef} or slower (\LFFn) (Fig.~\ref{fig:data}). In
larger data sets based on one-to-one communication, even broader
degree distributions have been
observed~\citep{LambiotteMobile,mobile2,mobile_DPLN}.  The NEMs give
rise to degree distributions that match these empirical data on large
acquaintance networks better than the nodal attribute models.

\subsection{\label{sec:results_c}Clustering spectrum}
\begin{figure}[htb]
\centering
\includegraphics[width=1.0\linewidth]{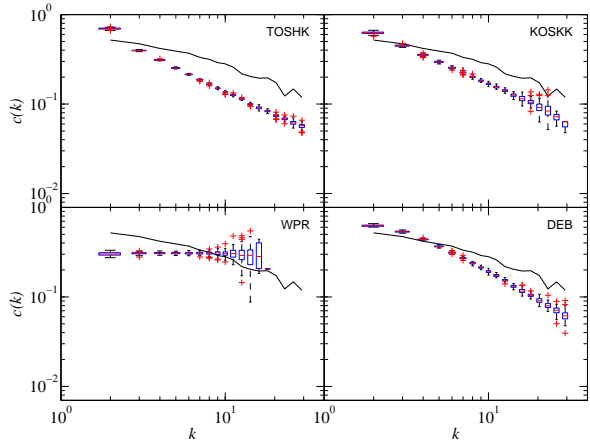}
\caption{Clustering spectrum $c(k)$ in the \LFF data (solid line) and
 in models fitted to it. Averaged over $100$ network realizations.}
\label{fig:clustering}
\end{figure}

Many network models display roughly an inverse relation between node
degree and clustering\footnote{This follows naturally in any model
where an increase in the number of links of a node goes hand in hand
with an increase in the number of triangles around it. If on average
increasing the degree $k$ of a node by one is accompanied by an
increase of the number $N_{\Delta}$ of triangles around the node by
$a$, the resulting clustering coefficient for a node of degree $k$
will be on average $c(k) = \frac{N_{\Delta}}{k (k-1) /2} = \frac{ak}{k
(k-1) /2} \approx \frac{2ak}{k^2} = \frac{2a}{k}$.}: $c(k) \sim
\frac{1}{k}$.  This holds true also for most of the NEMs studied here,
of which TOSHK, KOSKK, and DEB are shown in Fig.~\ref{fig:clustering},
as well as for the ERGM1 model (not shown). The figures display fits
to \LFF data, but results are similar for the \email fits.  In
contrast, the homophily mechanism on which the nodal attribute models
are based is seen to produce a flat clustering spectrum $c(k)=const$,
shown in Fig.~\ref{fig:clustering} for the \LFF fit of the WPR model.
In all empirical network data that we have come across, including both
of our data sets (Fig.~\ref{fig:clustering}) as well as acquaintance
networks based on Messenger and mobile phone calls
(e.g. \citet{mobile2}, \citet{MSnetwork}), clustering $c(k)$ decreases
with increasing degree $k$ of a node.  This indicates that attribute
based homophily alone does not seem to explain observed network
structures, supporting the findings by \citet{vip} and
\citet{HunterGoodnessOfFit}.

\subsection{\label{sec:results_geodesic}Geodesic paths}
\begin{figure}[htb]
\centering
\includegraphics[width=1.0\linewidth]{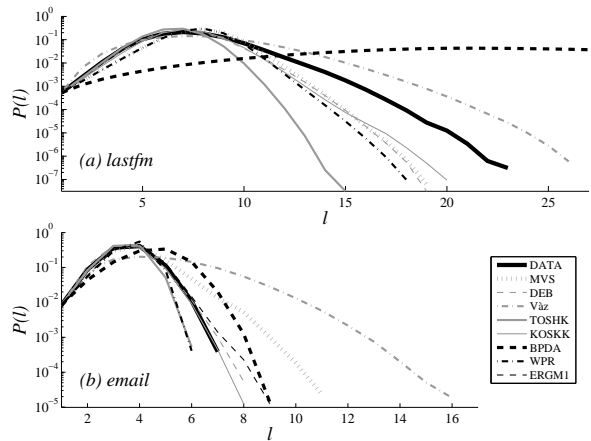}
\caption{Distributions $P(l)$ of geodesic path lengths $l$ in models
fitted to (a) the \LFF data, and (b) the \email data. The data is
shown as a thick black line in each panel. Averaged over $100$ network realizations.}
\label{fig:paths}
\end{figure}

Apart from the nodal attribute model treated one-dimensionally (BPDA),
in which average geodesic path lengths are strikingly long compared to
the data, all networks display reasonable path length distributions
(Fig.~\ref{fig:paths}). The dynamical NEMs and the TOSHK model are
slightly too compact, with largest path lengths falling below those in
the data.  The V\'az model, surprisingly, has rather long geodesic
paths despite its broad degree distribution. Generally, high degree
nodes decrease path lengths across the network, but the high
assortativity of the V\'az networks seems to counter the effect. For
reference, even in an extremely large acquaintance network of several
million individuals worldwide~\citep{MSnetwork}, the average distance
between two individuals is $6.6$, and path lengths up to $29$ have
been found.

\subsection{\label{sec:results_communities}Community structure}
\begin{figure}[htb]
\centering
\includegraphics[width=1.0\linewidth]{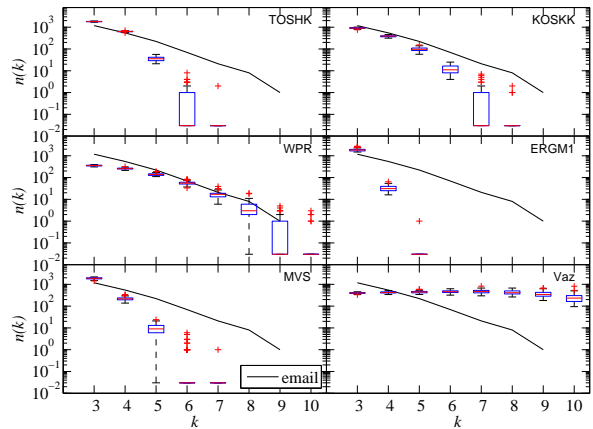}
\caption{ Number $n(k)$ of cliques of size $k$ in the model networks
fitted to the \email data, shown as a solid line in each panel for
reference.  Cliques within larger cliques, such as triangles within a
$4$-clique, are not counted. Averaged over $100$ network realizations.}
\label{fig:cliques}
\end{figure}

\begin{figure}[htb]
\centering
\includegraphics[width=0.8\linewidth]{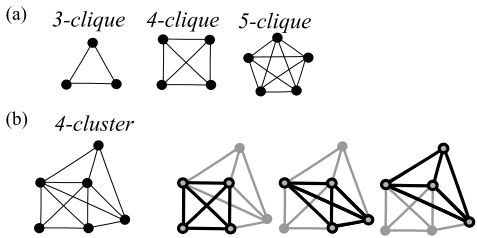}
\caption{(a) $k$-cliques for $k=3,4,5$. (b) An example of a $4$-cluster with $6$
nodes, highlighting the $4$-cliques from which it is formed.}
\label{fig:schematic_cliques}
\end{figure}

\paragraph{Cliques} Finally, we assess the community structure
of the networks.  Perhaps the simplest possible measure of community
structure is the number of cliques
(Fig.~\ref{fig:schematic_cliques}a), or fully connected subgraphs, of
different sizes. Figure~\ref{fig:cliques} displays the average numbers
of cliques in the model networks. Because each network has roughly an
equal number of nodes and links, the different numbers of cliques are
due to the arrangement of links within the network and not to
differences in global link density. It turns out that the NAMs produce
clique size distributions that match the data sets fairly well in both
fits. The WPR model, fitted to the \email data, is shown in
Fig.~\ref{fig:cliques}.  The KOSKK and DEB models also produce
distributions roughly comparable to the empirical data, and the \vaz
model in fact produces far too many large cliques when link density is
high (Fig.~\ref{fig:cliques}). The MVS and TOSHK models have trouble
producing large enough cliques when link density is low (the \LFF
fits). A possible explanation of why the MVS model produces very few
cliques is indicated by the comparison of Section~\ref{sec:fourfield},
where node deletion is seen to preserve more cliques than link
deletion.  The parametrization of the TOSHK model, requiring that the
number of secondary contacts be drawn from a uniform distribution,
severely limits the number of coincident triangles and hence cliques
which can be formed. The ERGM1 model produces the fewest cliques of
all the models.

\begin{figure}[htb]
\centering
\includegraphics[width=1.0\linewidth]{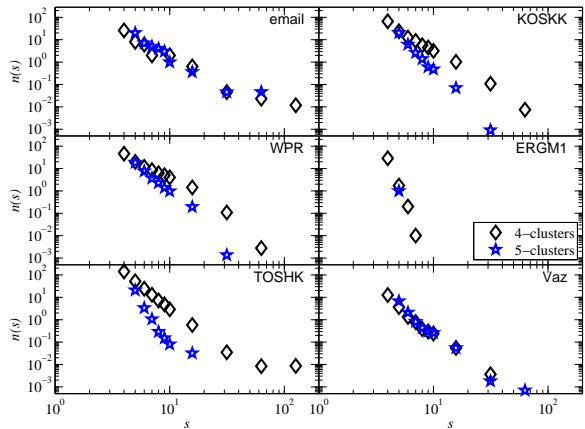}
  \caption{Average number $n(s)$ of $k$-clusters of size $s$ in a  
  network, for $k=4$ ({\Large$\diamond$}) and $k=5$ ({\Large$\star$}),
  in the \email data, and in models fitted to it.
Averaged over $100$ network realizations. }
  \label{fig:k-clusters}
\end{figure}

\paragraph{\label{sec:results_kclusters}k-clusters} 
We also identify communities using the {\em k-clique-percolation}
method developed by~\citet{kclique}. The method defines a {\em
k-cluster} as a subgraph within which all nodes can be reached by
'rolling' a $k$-clique such that all except one of its nodes are fixed
(see Fig.~\ref{fig:schematic_cliques}b).  Figure~\ref{fig:k-clusters}
displays the size distributions of $k$-clusters with $k=4$ and $k=5$
for several models fitted to the \email data. As the ERGM1 model
produced very few cliques apart from triangles, it cannot generate large
$k$-clusters for $k>3$. The other models generally produce $4$-cluster size
distributions roughly matching the data, but large $5$-clusters are
relatively few. The V\'az model generates networks containing very
large $k$-clusters with high values of $k$. These are likely due to an
extremely dense 'core' formed around nodes that joined the network
early on. For example, each of the $100$ network realizations
contained $10$-cluster of size $s=72 \pm 15$ (not shown). Such dense
clusters are not generally observed in empirical data. For example, in
the \LFF and \email data sets, the largest $10$-clusters are of sizes
$10$ and $12$, respectively.

\paragraph{\label{results_overlap}Role of links with low overlap}
In both of our empirical networks, as well as in the networks
generated by the studied models, a rather large fraction of edges does
not participate in any triangles. In the \LFF and \email data, the
fraction of such edges is $31.2 \%$ and $22.4\%$
respectively\footnote{This might be due to the nature of our empirical
data sets, which are sampled from networks that are constantly growing
with links and nodes accumulating over time. In them, a relatively
large fraction of nodes are newcomers who have only established a few
links to the system, such that triangles have not yet been formed
around them.}.  The DEB, TOSHK, \vazn, and ERGM1 models produce
slightly too few such links ($20$ to $22\%$ in the the \LFF fits and
$4$ to $5\%$ in the \email fits, except $12,6\%$ in the \email fit of
ERGM1), whereas the nodal attribute models and KOSKK tend to generate
slightly too many of them ($35$ to $40\%$ in the the \LFF fits and
$27$ to $41\%$ in the \email fits).
% In the \email fit of ERGM1,  $12.6\%$ of links do not participate in triangles. 

\begin{figure}[htb]
\centering
\includegraphics[width=0.75\linewidth]{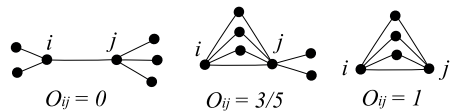}
\caption{Overlap $O_{ij}$ of edge $e_{ij}$.}
\label{fig:overlap}
\end{figure}

We can ask what structural role is played by links that do not
participate in triangles, or more generally, by links whose end nodes
share only a small fraction of their neighbors. Within a community,
adjacent nodes tend to share many neighbors, while for edges between
communities, the neighborhoods of the end nodes will not overlap
much. This can be quantified using a measure called the {\em overlap}
$O_{ij}$ \citep{mobile2}, which could be interpreted as a modification
of the edgewise shared partners measure~\citep{curvedERGM}, measuring
the {\em fraction} instead of the {\em number} of edgewise shared
partners for the end nodes of an edge. The measure also bears
resemblance to the Jaccard coefficient~\citep{Jaccard}. The overlap
is defined as $O_{ij}=\frac{n_{ij}}{(k_i-1) + (k_j -1) - n_{ij}}$,
where $n_{ij}$ is the number of neighbors common to both nodes $i$ and
$j$, and $k_i$ and $k_j$ are their degrees (Fig.~\ref{fig:overlap}).

\begin{figure}[htb]
\centering
\includegraphics[width=0.85\linewidth]{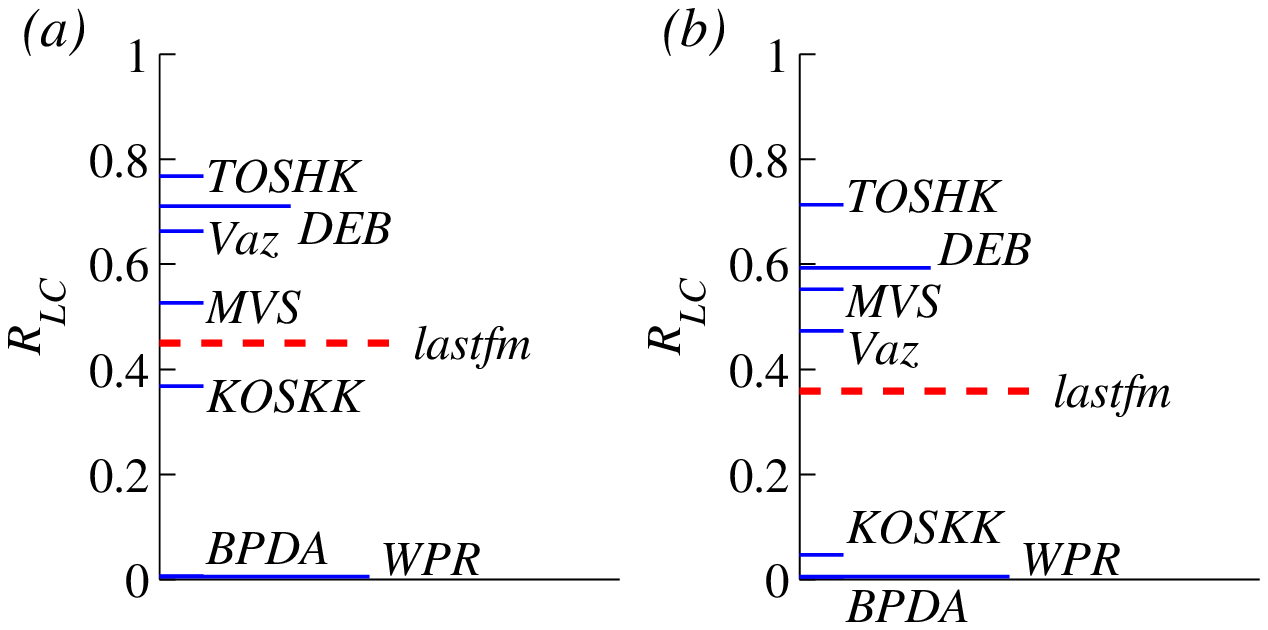}\\
\includegraphics[width=0.7\linewidth]{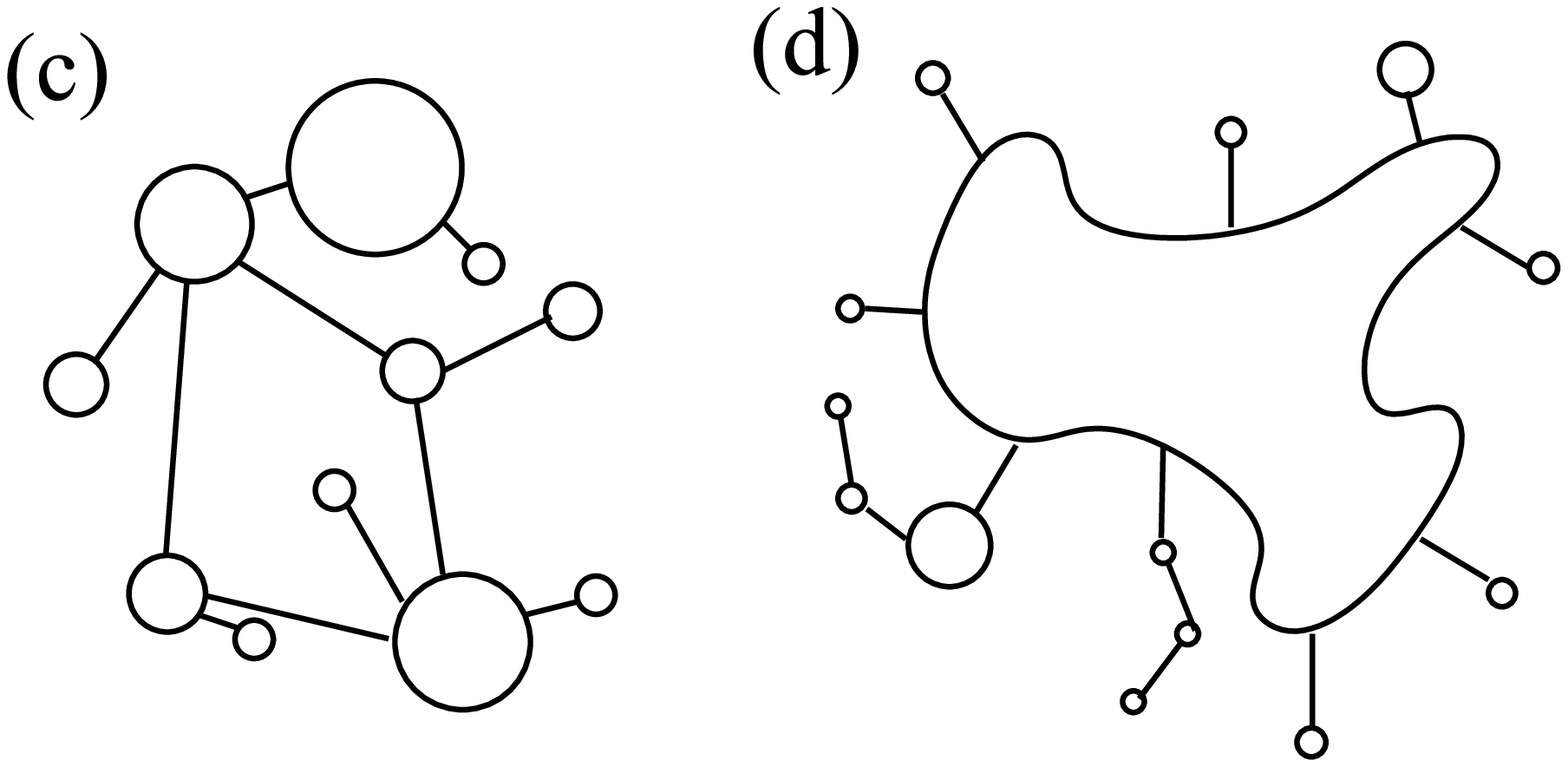}
\caption{ (a) Relative size $R_{LC}$ of the largest connected
component in the models fitted to the \LFF data after removing links
with overlap $O=0$.  (b) To show that the breakdown of the nodal
attribute models was not simply due to a larger number of links
removed, we now remove the same fraction of the lowest overlap links
from all models and data ($41\%$, the maximum fraction removed in
Fig.~\ref{fig:NLC_thresholded}(a)). Data averaged over $100$ network
realizations.  (c and d) Schematic depiction of the structural
differences related to links with low overlap (links whose end nodes
share only a small fraction of their neighbors).  (c) Low overlap
links connect small, relatively tightly bound clusters together.  (d)
The network contains a core that does not disintegrate when low
overlap links are removed.  }
\label{fig:NLC_thresholded}
\end{figure}

Removing low-overlap-links will separate dense, loosely interconnected
communities from one another. This turns out to discern the nodal
attribute models and the KOSKK model from the other models and our
empirical data.  Figure~\ref{fig:NLC_thresholded}(a) displays the 
relative sizes of the largest component after removing links that do not
participate in triangles for the \LFF data and the models fitted to
it. The nodal attribute models break down to small clusters, whereas
in the other models a large core remains.

As noted earlier, the NAMs contain more zero-overlap links than the
other models. Hence, it is useful to check whether their breakdown was
due to a larger fraction of removed links.  We can test this by
removing an equal fraction of links from all networks ($41 \%$,
the maximum fraction of links removed from any network when only
non-triangle-links were removed) (Fig.~\ref{fig:NLC_thresholded}b). We
remove links in increasing order of overlap $O_{ij}$.  Again, a core
remains intact in most of the NEMs, whereas the NAMs and the KOSKK
network break down, indicating in these models the absence of a core,
and the role of low overlap links as bridges between clusters.

The link densities of the remaining components, $d=2 \; l /s(s-1)$, where $s$ is
the number of nodes in the component and $l$ the number of links, are
moreover observed to be slightly higher in the NAMs than in the other
models, despite the fact that more links were removed from them (not
shown). The above findings show that these networks consist of very
clear communities that are loosely interconnected. The other NEMs and
ERGM1 on the other hand contain a core that does not consist of such
loosely connected clusters. This difference is depicted schematically
in Fig.~\ref{fig:NLC_thresholded}(c,d).  

In the \email fits, link density in the network is higher, and for all
networks slightly larger overlap links need to be removed in order to
decompose them to small clusters (not shown), but the general
difference between the NEMs and NAMs remains. As the ERGM1 model was
only fitted to the \email data, it is not displayed in
Fig.~\ref{fig:NLC_thresholded}. Removing low overlap links did not
reduce the largest component of the ERGM1 networks practically at all
- even after removing $50$ percent of links beginning with lowest
overlap, a core containing on average $93.6$ percent of the nodes
remains intact - consistently with the finding that the networks did
not contain many denser substructures such as cliques or $k$-clusters.

\subsection{\label{sec:fourfield}Differences in network structure
  resulting from choice of mechanisms for triadic closure and link deletion} 
\begin{figure}[htb]
\centering
\includegraphics[width=1.0\linewidth]{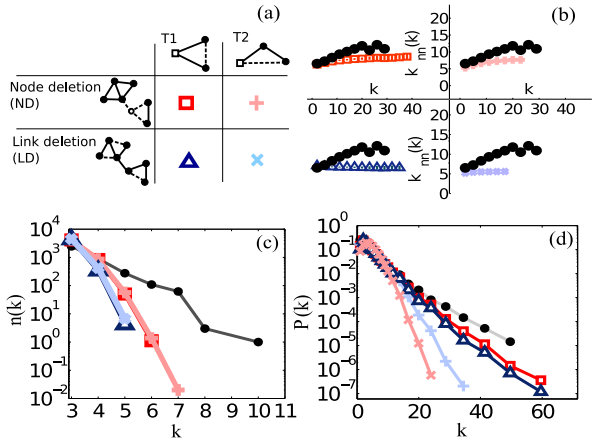}
\caption{Comparison of mechanisms employed in dynamical network
    evolution models. (a) Two mechanisms of triadic closure (T1 and
    T2) are combined with two ways of deleting links (node deletion
    refers to deleting all links of a node, and link deletion refers
    to deleting randomly selected links). The same symbols are used in
    panels (b)-(d). (b) Average nearest neighbor degree \knnave with
    respect to node degree $k$, variants arranged as in the schematic
    figure. The \LFF data is also shown in each panel.  (c) Number
    $n(k)$ of cliques of each size $k$. Smaller cliques within larger
    cliques are not counted.  (d) Degree distribution $P(k)$. Averaged
    over $100$ network realizations.}
\label{fig:DEBvariants}
\end{figure}

Here, we will examine the differences in network structure resulting
from combinations of the mechanisms of link generation (T1,T2) and
deletion (ND,LD) emplyed in dynamical network evolution models.
Taking as a starting point the simplest of the dynamical models (DEB),
in which a newcomer will link to exactly two uniformly randomly chosen
nodes, after which it will only initiate triadic closure steps, we
study all four combinations of the mechanisms
(Fig.~\ref{fig:DEBvariants}, a). Two findings speak in favor of using
the node deletion mechanism: The model variants using T1 show a
clearly assortative relation, suitable for social network models,
whereas the T2 networks are dissortative or very weakly assortative
(Fig.~\ref{fig:DEBvariants}, b). Node deletion also preserves more
cliques in the network, a desirable feature for social networks
(Fig.~\ref{fig:DEBvariants}, c). The larger number of cliques
preserved by node deletion is not explained by the clustering
coefficients, which turned out to be similar in all networks.  The
parameters were selected such that \nl and \kave matched the \LFF
data.

The choice of triangle generation mechanism, on the other hand, is
seen to affect the degree distribution. Networks generated with the T1
mechanism have higher degree nodes than those using the T2 mechanism
(Fig.~\ref{fig:DEBvariants}, d). This is because following a link is
more likely to lead to a high degree node than picking a node
randomly. Because in T1 both of the nodes gaining a link in the triad
formation step are chosen by following a link, high degree nodes
obtain more additional links than when the T2 mechanism is used, in
which one of the nodes is chosen randomly. The choice of T1 or T2 does
not seem to have an effect on the number or size of cliques generated,
nor on degree-degree correlations. 

\vspace{-12pt}
\section{\label{sec:discussion}Summary and discussion}
In order to assess the resemblance to empirical networks of the many
models for social networks that have been published in recent years in
the physics-oriented complex networks literature, we have fitted these
models to empirical data and assessed their structure.  We have also
compared these models with an exponential random graph model that
incorporates recently proposed specifications, in the first systematic
comparison between models from these families. In addition to
comparing structural features of networks produced by the models, we
have discussed the different philosophies underlying the model types.

The structural features we focused on are similar in the two included
empirical data sets as in numerous other large empirical social
networks~\citep{mobile2,mobile_DPLN,MSnetwork} in that they have
highly skewed degree distributions, high average clustering
coefficients, decreasing clustering spectra $c(k)$, and positive
degree-degree-correlations $r$.  Therefore, any widely applicable
model for social networks should be able to approximately reproduce
the average values and distributions of their main characteristic
features. However, as the philosophy behind the NEMs studied here is
to explain the emergence of common structural features of social
networks, we shouldn't expect them to capture perfectly all features
of particular empirical data sets. Our main motivation for fitting the
models to the selected target features was to unify approximately some of
their properties, in order to compare meaningfully their higher order
properties, such as the degree distribution and community structure.
These are not likely to be drastically altered by small differences in
the average values. Hence, we do not consider an accurate fit in the
average quantities of extreme importance.  

For almost all models, we saw that average largest component size
\nlave and average degrees \kave could be fitted closely to both
empirical data sets. In the ERGM1 model, we compromized matching
average degree in order to obtain a reasonable clustering coefficient.
Adaptability was limited by the number of free parameters. The models
DEB and \vazn, which had only one free parameter in addition to
network size, turned out to have excessively high average clustering
coefficients even for the moderate average degrees displayed by our
two data sets. For most of the other models, clustering could be tuned
rather closely. Being able to match the targeted average values of
these two data sets does not guarantee that a model is able to match
those features in other empirical data, however. In this sense, the
generalisability of conclusions based on only two data sets is
limited. 

\newcommand{\datawidth}{2.3cm}
\newcommand{\NAMwidth}{1.8cm}
\newcommand{\gNEMwidth}{2.6cm}
\newcommand{\dNEMwidth}{2.6cm}
\newcommand{\s}{\smallskip }
\newcommand{\sm}{\smallskip \smallskip }

\begin{table*}[htb]
\caption{Summary of structural properties of networks generated with the studied models.  \s
}
\tiny
\begin{tabularx}{1.0\linewidth}{ |X | l | l | l | l | l |}
\hline
\parbox{\paramswidth}{ \s Property \s }  & \LFF and \email &  NAMs & dynamical NEMs & growing NEMs  & ERGM1   \\ 
\hline
\parbox{3 cm}{ \sm degree distribution \sm} & relatively broad & peaked & relatively broad & broad & relatively broad  \\  
\parbox{3 cm}{\sm clustering spectrum \sm} & decreasing  & flat  & decreasing & decreasing & decreasing  \\
\parbox{3 cm}{\sm assortativity \sm  } & yes  & yes (high)  & yes (weak) & yes (moderate/high) & yes (weak)  \\
\parbox{3 cm}{\sm geodesic path lengths \s   } & -  & \parbox{\NAMwidth}{ in 1D,
  too long \\ longest paths }  & reasonable &
  \parbox{\gNEMwidth}{ reasonable  } &  reasonable \\
\parbox{3 cm}{\sm  cliques \sm } & many large cliques & many large cliques &\parbox{\dNEMwidth}{many in KOSKK, \\ fewest in MVS}  & 
\parbox{\paramswidth}{ too few  in TOSHK, \\ exceedingly in \vaz } & very few \\
\parbox{3 cm}{\sm  $k$-clusters \sm } & \parbox{\datawidth}{many large $k$-clusters \\ for $k=4$ and $k=5$}  & reasonable  &  \parbox{\dNEMwidth}{reasonable in DEB and
\\ KOSKK,  too few in MVS }& \parbox{\gNEMwidth}{ in \vazn, exceedingly
  large \\
  $k$-clusters  with large $k$ } & no large $k$-clusters   \\ 
\parbox{3 cm}{\sm  consisting of dense \\ clusters interconnected\\ by low-overlap links \sm } & no  & yes  &
\parbox{\dNEMwidth}{yes (KOSKK),  \\no  (DEB and MVS) }& \parbox{\gNEMwidth}{ no } & no  \\ 
\hline
\end{tabularx} 
\label{table:results} 
\end{table*} 

Table~\ref{table:results} summarizes the structural features in
networks resulting from the different model types. Nodal attribute
models (NAMs) in which the nodes are located with uniform probability
in the underlying social space and links are based solely on
homophily, produce a clustering spectrum $c(k)$ strikingly different
from observed data, indicating that it is not a sufficient description
of the mechanisms at play in the formation of social networks. They
also produce peaked degree distributions without very high degree
nodes that do no agree with empirical data on large scale social
networks. The homophily principle employed in the nodal attribute
models is seen to be sufficient for producing strong positive
degree-degree correlations. This is a direct result of the dependence
of link probability on distance: because high degree nodes appear in
locations with a dense population of nodes, their neighbors will also
tend to have high degree. The NAMs also generate networks containing a
large number of cliques and consisting of dense clusters loosely
connected with low overlap links. Their clustered structure appears
more pronounced than in the data.

We find that many of the studied network evolution models (NEMs)
produce broader degree distributions and decreasing clustering spectra
that agree more closely with empirical data. Most of them also
generate assortative networks, although typically not to the same
extent as in the data, and many large cliques and $k$-clusters. In
the dynamical NEMs, node deletion is seen to produce more assortative
networks than link deletion. With respect to thresholding by overlap,
the dynamical KOSKK model displayed the clearest clustered structure
of all the NEMs. This shows that the weights employed in tie formation
in the KOSKK model play an important role in the formation of
community structure, as the authors observed~\citep{KOSKK}. The other
NEMs produced networks which, in accordance with the data, contained a
large core that did not break apart when low overlap links were
removed.

The exponential random graph model ERGM1 incorporating recently
proposed terms for structural
dependencies~\citep{newSpecifications,HunterGoodnessOfFit,ERGMRecentDevelopments}
was seen to generate very few large cliques. It did produce
assortative networks, although with relatively low
assortativity. These terms had earlier been employed without
difficulty when fitting ERGMs to a large social
network~\citep{GoodreauERGMFittedToSchoolData}. However, we
encountered problems of multimodality with the model.

Very large social networks of millions of individuals, within a
country or worldwide, can be assessed with data provided by modern
electronic communications, such as mobile phone calls~\citep{mobile1}
or instant messaging~\citep{MSnetwork}. The data have revealed
features of large scale networks of human interaction that could not
be discerned from a small subnetwork. These include the tails of
highly skewed distributions as well as distributions of mesoscale
structures, such as the size distribution of communities. Modeling
the structure observed in large networks benefits from the ability to
generate networks of comparable size. NEMs and NAMs fulfill this
requirement.

Using realistic models for social networks in simulation studies of
social processes is essential in light of the knowledge that network
structure influences many processes~\citep{socialDynamicsCastellano},
such as the emergence and survival of cooperation~\citep{lozano},
spreading of information~\citep{mobile1,rumor} or
epidemics~\citep{spreadingInCorrelatedNetworks}, and coexistence of
opinions~\citep{Lambiotte:2007oq}.

Many structural characteristics of social networks were attained even
with very simple mechanisms. However, neither the nodal attribute
models based on homophily, nor the network evolution models based on
\tfcn, were able to reproduce all important features of social
networks. As both mechanisms obviously are present in the evolution
social networks, a combination of the model types could yield more
realistic network models.

\section*{\label{sec:acknowledgements}Acknowledgements}
We acknowledge the Academy of Finland, the Finnish Center of
Excellence Program 2006-2011, Project No. 213470. R.T. is supported by
the ComMIT graduate school.

{\small
\appendix
\section{Appendix}

\subsection{\label{sec:characteristics}Basic network measures}

The network representation of social contacts consists of {\it nodes}
representing the individuals, and {\it links} representing the ties
between them. An overline is used to denote averaging over all nodes
(or links) within the network, or across several networks. We denote
by $N$ the number of nodes in a network, i.e. network \em{size}.  A
\em{component} of a network is a connected subset of nodes.  In this
paper, we study the {\it largest component} LC of each network. We
denote its size by $N_{LC}$.  The number of network neighbors of a
node is called its {\em degree} $k$. An isolated node has degree zero.

A measure of local triangle density, the {\em clustering coefficient},
describes the extent to which the neighbors of node $i$ are acquainted
with one another: if none on them know each other, $c_i$ is zero,
while if all of them are acquainted, $c_i=1$.  For a node $i$ with
degree $k_i$ and belonging to $T_i$ triangles, the clustering
coefficient is defined as
\begin{equation}
c_i= \frac{T_i}{k_i ( k_i-1) /2},
\label{eq:clustering}
\end{equation}
where the denominator ${k_i ( k_i-1) /2}$ expresses the maximum
possible number of triangles $i$ could belong to given its degree.
The clustering coefficient is not defined for nodes with degree $k<2$.
The average clustering coefficient, averaged over all nodes with
$k\ge2$ in the network, is denoted \caven.  $c(k)$ denotes
the average clustering coefficient of nodes with degree $k$. The
curve $c(k)$ is called the {\it clustering spectrum}.
 
In large empirical social networks, typically high degree nodes tend
to be linked to other high-degree nodes, and low-degree nodes tend to
be linked among themselves. One way of quantifying this effect is
using linear correlation, or the Pearson correlation coefficient,
between the degrees $k_i$ and $k_j$ of pairs of connected nodes. This
is also called the \emph{assortativity coefficient}
$r$~\citep{NewmanAssortative}:
\begin{displaymath} 
r 
= \frac{ \sum_{e} k_i k_j / E -   \left[  \sum_{e} \frac{1}{2} (k_i +
    k_j) \right] ^2 / E^2} {\sum_{e} \frac{1}{2} (k_i^2 + k_j^2) /E -
  \left[  \sum_{e} \frac{1}{2} (k_i + k_j) \right] ^2 /E^2 },
\end{displaymath}
where $E$ is the total number of links in the network. Assortativity
can also be quantified using the measure {\em average nearest neighbor
degree} $\overline{k_{nn}}(k)$, found by taking all nodes with degree
$k$, and averaging the degrees of their neighbors. If the curve
$\overline{k_{nn}}(k)$ plotted against $k$ has a positive trend, nodes
with high degree typically also have high-degree neighbors, and hence
the network is assortative.

The geodesic path length $l_{ij}$ between nodes $i$ and $j$ in a
network means the minimum number of links that need to be traversed in
order to get from $i$ to $j$. The average length \lave of geodesic
paths between nodes describes the compactness of the network.

\subsection{\label{sec:appendix_fitting}Determining optimal network parameters}
Our fitting method consists of simulating network realizations with
different values of the model parameters, and finding the values
(points in the \emph{parameter space}) that produce the best match to
the following features of the empirical data sets: average degree
\kaven, average clustering coefficient \caven, and average geodesic
path lengths \lave (in this order of importance, depending on the
number of model parameters).  This approach deviates from the
tradition of maximum likelihood estimation for fitting probabilistic
models.

We attempt to minimize the relative error in each chosen feature. For
example, for average degree \kave in a model with given parameter
values $\textbf{p}$, being fitted to a data set with average degree
$\overline{k}^{target}$, the relative error is
\begin{equation}
  | {\bf \epsilon}_{ \overline{k} (\textbf{p})} | = \left| \frac{ \overline{k} (\textbf{p}) - \overline{k}^{target} }{ \overline{k}^{target} } \right| .
\label{eq:errorfunction}
\end{equation}
The errors for each feature are combined in the \emph{error function}
$f(\textbf{p})$, whose norm $|f(\textbf{p})| $ is minimized. For example, if fitting to
$N_{LC}$, \kave  and \caven, the error function and its norm take the shape
\begin{equation}
f(\textbf{p}) = \left[ w_{N_{LC}} \epsilon_{N_{LC} }  \quad
  w_{\bar{k}} \epsilon_{\bar{k}} \quad
  w_{\bar{c}} \epsilon_{\bar{c}} \right],
\label{eq:targetfunction}
\end{equation}
 and its norm 
\begin{equation}
|f(\textbf{p})| = \sqrt{  w_{N_{LC}}^2 \epsilon_{N_{LC}}^2  + 
  w_{\bar{k}}^2 \epsilon_{\bar{k}}^2  +
  w_{\bar{c}}^2 \epsilon_{\bar{c}}^2. 
} 
\label{eq:targetfunction_norm}
\end{equation}
The error function should have equally many components as there are
network parameters.  We chose weights that reflected the order of
importance given to the targeted features, putting the most emphasis
on matching the number of nodes and links, less on clustering, and
least on average geodesic path lengths. It turned out that for nearly
all of the models (DEB, MVS, KOSKK, Vaz, BPDA, \email fit of WPR) the
result was insensitive to weights, because the models were able to
match the target values closely (up to the number of model
parameters).
In optimizing the models DEB and KOSKK, we used a linear approximation
for the components of the error function, iteratively refining the
approximation close to the optimum. For MVS, we used the the well
established {\it Nelder-Mead method}~\citep{neldermead}, which
involves calculating values of the error function at the corners of a
simplex (a triangle in 2-dimensional space, a tetrahedron in 3D). The
optimal value of the error function is iteratively approached by
rolling one corner of the simplex over the others such that the object
moves towards the region where the function gets optimal values. The
diameter of the simplex is adjusted during iteration to increase
accuracy.

Optimization algorithms were not needed for the \vaz and BPDA models
and the \email fit of WPR. For the \vaz model, a very good
approximation for the optimal value of $u$ can be obtained
analytically. This estimate could be refined manually. For the BPDA
model, the analytical estimates for \kave and \cave derived by the
authors could be used as a starting point in optimization.  We refined
the initial estimates by first adjusting $\alpha$ to find the correct
value for the clustering coefficient, and then changing $b$ until the
correct mean degree was found. For small enough adjustments, the
latter corrections did not affect the value of \caven. The adjustments
were done by trial and error, but it was not difficult to get an
accurate fit for mean degree and clustering in this manner.  For the
\email fit of WPR, it turned out that $N_{LC} \approx N$, and hence
the number of free parameters was reduced.  $p$ was set to obtain
desired average degree, and the two remaining parameters were
optimized by generating networks with a grid of their values.

Exact fits could not be obtained for TOSHK, ERGM1, and the \LFF fit of
WPR.  For WPR, we used weights $[w_{N_{LC}}, w_k, w_c, w_l] = [4\; 4\;
2\; 1]$ and grid optimization similarly as in the \email case,
although it was costly in four dimensions. Obtaining values in a grid
enabled us to visulize the dependence of the targeted features on the
model parameters. It turned out that assortativity and clustering
varied closely together, rendering assortativity useless as a fitting
target if clustering was used; hence we used average geodesic path
lengths, which enabled an optimum to be determined.  As the TOSHK
model has only one continuous parameter $p$, it suffices to optimize
$p$ for all values of the discrete parameter $k$ below some
$k_{\max}$, making sure that $k_{\max}$ is large enough. The parameter
$p$ was optimized to reach the desired mean degree for each $k$, and
the pair $(k,p^{\textrm{opt}}(k))$ that provided the best match to the
desired \cave was selected as the optimum. Optimization was carried
out with the Matlab optimization toolbox function \emph{fminbnd.m},
which is based on golden section search and parabolic interpolation.

For the remaining case in which no exact match was found (ERGM1), we
attempted using the linear approximation method and Nelder-Mead
algorithm described above, as well as other, potentially more robust
methods~\citep{elsterNeumaier,snobfit}, but these failed likely due to
multimodality of the probability
distribution. Figure~\ref{fig:ERGM_transition} illustrates the
instability we encountered when attempting to fit the ERGM1 model to
the \email data.  The panels display average degree \kave (a) and
average clustering coefficient \cave (b) in networks generated with
various values of $\theta_L$, with the other parameters kept constant
at the values listed in Table~\ref{table:targeted_features}. For each
value of $\theta_L$, $60$ network realizations are shown (drawn from
MCMC chains with burn-in $5 \times 10^7$ steps, and 5 realizations
taken from each chain at intervals of $10^7$).  Because $\theta_L$
controls the number of random links, an increase in $\theta_L$
generally increases average degree and decreases average clustering.
However, at roughly $\theta_L =-6.961$ we observe a sudden transition
into a much denser, less clustered network.
\begin{figure}[htb]
\centering
\includegraphics[width=0.475\linewidth]{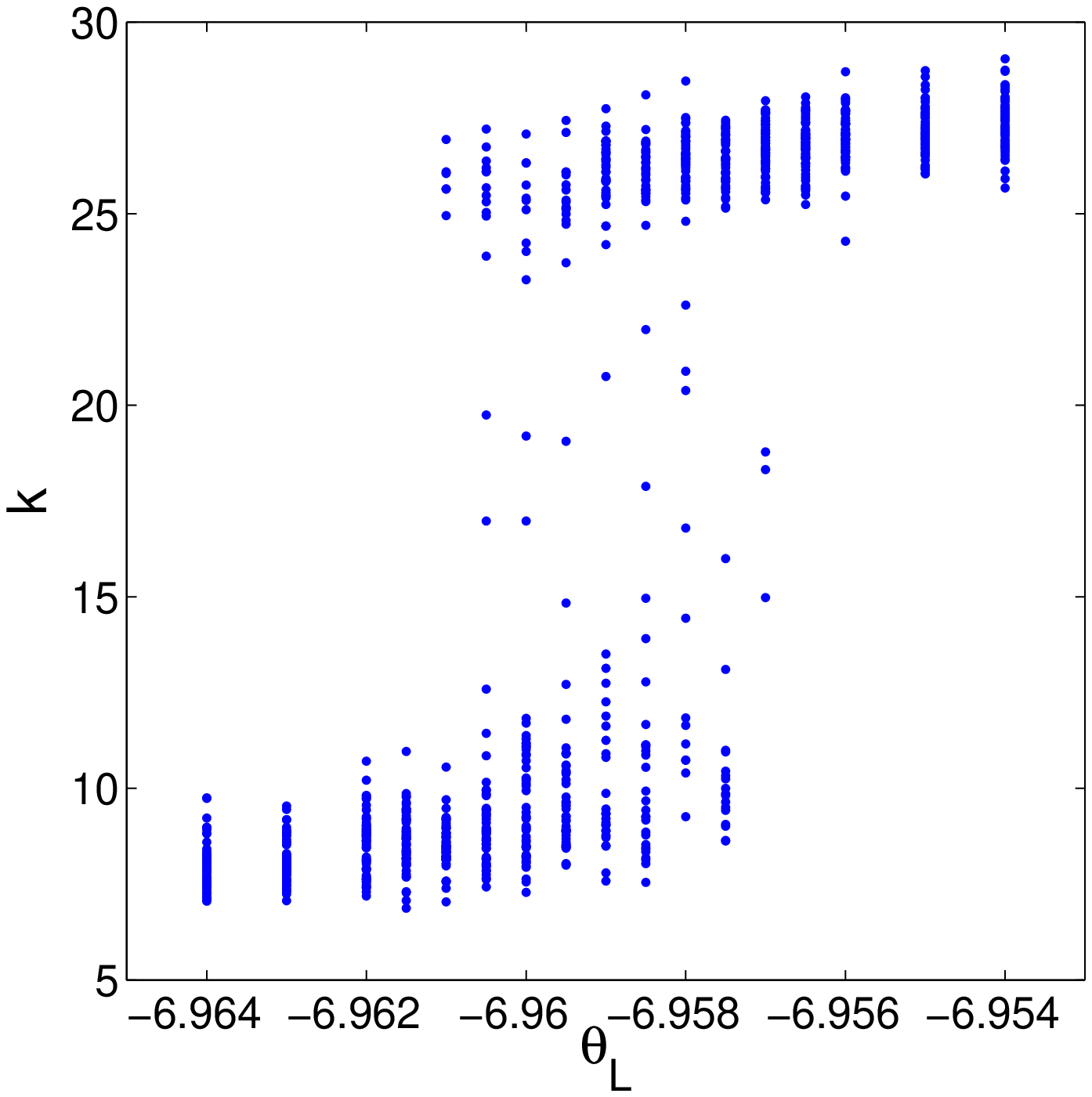}
\includegraphics[width=0.49\linewidth]{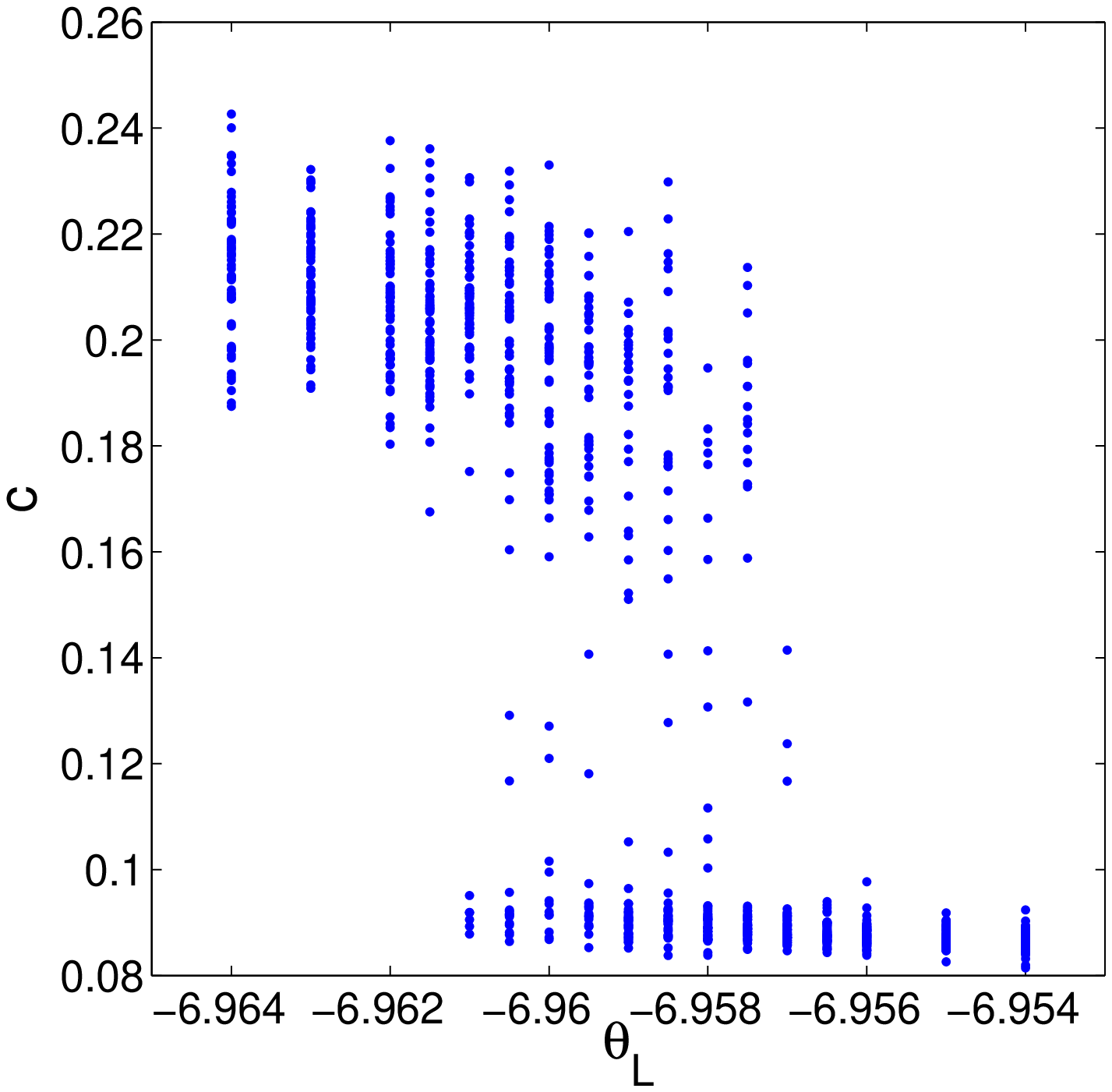}
\caption{(a) Average degree \kave and (b) average clustering coefficient
  \cave in networks generated from the model ERGM1  with different
  values of $\theta_L$.}
\label{fig:ERGM_transition}
\end{figure}

}

\bibliographystyle{elsarticle-harv}
\bibliography{toivonen}
% \bibliography{refs}

\end{document}